\documentclass[journal=jacsat,manuscript=article]{achemso}
\DeclareUnicodeCharacter{202F}{\,}
\usepackage[version=3]{mhchem} 
\usepackage[hidelinks]{hyperref}
\usepackage{textcomp}
\usepackage{color}
\graphicspath{{./Images/}}
\usepackage{mathtools}
\usepackage{graphicx}
\usepackage[dvipsnames]{xcolor}
\usepackage{ulem}
\usepackage{pdfpages}
\usepackage{changes}
\usepackage{todonotes}

\makeatletter
\AtBeginDocument{\let\LS@rot\@undefined}
\makeatother

\usepackage{pgffor}
\author{Enise Kartal}
\email{e.kartal@tudelft.nl}
\affiliation{Department of Precision and Microsystems Engineering, Delft University of Technology, Delft, The Netherlands}

\author{Oriel Shoshani}
\affiliation{Department of Mechanical Engineering, Ben-Gurion University, Beer-Sheva, Israel}

\author{Elena Botnaru}
\affiliation{Department of Precision and Microsystems Engineering, Delft University of Technology, Delft, The Netherlands}

\author{Alberto Martín-Pérez}
\affiliation{Department of Precision and Microsystems Engineering, Delft University of Technology, Delft, The Netherlands}

\author{Tomás Manzaneque}
\affiliation{Department of Microelectronics, Delft University of Technology, Delft, The Netherlands}

\author{Farbod Alijani}
\email{f.alijani@tudelft.nl}
\affiliation{Department of Precision and Microsystems Engineering, Delft University of Technology, Delft, The Netherlands}

\title[Short Title for Header]{
  {\fontsize{16pt}{20pt}\selectfont Frequency Stability of Graphene \\ Nonlinear Parametric Oscillator}\vspace{20pt}
}


\begin{document}






\begin{abstract}

High-frequency stability is crucial for the performance of graphene resonators in sensing and timekeeping applications. However, the extreme miniaturization and high mechanical compliance that make graphene attractive also render it highly susceptible to nonlinearities, degrading frequency stability. Here, we demonstrate that graphene parametric oscillators provide an alternative nonlinear operating regime, where short-term frequency stability can be enhanced despite strong nonlinearity. By operating graphene resonators in a phase-locked loop (PLL), we experimentally demonstrate that parametric oscillations in the post-bifurcation regime achieve lower Allan deviation at fast integration times than Duffing oscillations at identical amplitudes. This improvement originates from strong nonlinear damping inherent to parametric oscillators, which suppresses amplitude-to-frequency noise conversion at large amplitudes. A minimal theoretical model captures observed phase diffusion and identifies nonlinear damping as the dominant mechanism governing phase noise reduction. These results highlight the role of nonlinear dissipation in enabling precision sensing beyond conventional limits of graphene oscillators.

\end{abstract}

Nanomechanical resonators have emerged as excellent platforms for exploring rich dynamical phenomena inaccessible at the macroscale \cite{bachtold2022mesoscopic, belardinelli2025hidden,li2025cascade,arjmandi2025mechanical}, as well as for developing innovative nanodevices. Beyond their established role in sensing \cite{bachtold2012yoctogram,kaynak2023paddle,roslon2022probing}, they have also been explored for signal processing \cite{kartal2025reservoir} and timekeeping \cite{kenig2012passive,villanueva2013surpassing,matheny2014phase}. In these contexts, they are often integrated into a feedback loop that compensates for dissipation losses and sustains their periodic motion, forming oscillators.  A key performance metric of such oscillators is their frequency stability, which is commonly divided into short-term and long-term stability. The short-term stability is influenced by phase noise, jitter, and other stochastic effects, whereas long-term stability is limited by drifts arising from aging, temperature, and voltage variations \cite{ferre1997draft}. Macroscopic atomic clocks, which provide the ultimate time definitions and possess virtually zero drift in their energy level transition frequency \cite{audoin2001time,ludlow2015optical}, are associated with extraordinary long-term stability \cite{hinkley2013atomic,marshall2025high}, while MEMS and NEMS oscillators, which offer the integration and portability required for modern electronics, stand out in their short-term frequency stability\cite{kenig2012passive,villanueva2013surpassing,matheny2014phase,vig1999noise,bevsic2023schemes,shoshani2024extraordinary}.

Conventionally, MEMS and NEMS oscillators are operated in their linear regime to obtain predictable, single-valued responses. While maximizing the drive amplitude generally improves the signal-to-noise ratio (SNR) in these systems, this strategy is limited by the onset of nonlinear effects, where the effect of amplitude-to-frequency (A-F) noise conversion hinders the frequency stability \cite{zanette2012frequency,shoshani2024extraordinary}. This constraint is especially critical in applications where MEMS and NEMS oscillators function as resonant sensors, because extreme miniaturization reduces the linear dynamic range and enhances the sensitivity to both intrinsic and extrinsic noise sources \cite{sansa2016frequency}. For this reason, understanding and improving frequency stability in the nonlinear regime has become necessary to continue pushing the limits of nanomechanical systems \cite{kenig2012passive,villanueva2013surpassing,sansa2016frequency}. This is particularly true for atomically thin two-dimensional (2D) materials like graphene, where their ultralow mass and high mechanical compliance give rise to nonlinear dynamic phenomena even at small drive amplitudes \cite{steeneken2021dynamics}.  

In this work, we explore the frequency stability of graphene parametric oscillators operating in the nonlinear regime. Such systems are known to exhibit rich parametric resonances \cite{dolleman2018opto,keskekler2021tuning}. Yet, their implications for frequency stability remain incompletely understood. By operating graphene nanodrums in a phase-locked loop and driving them beyond their period-doubling bifurcation, we demonstrate that short-term frequency stability can be enhanced up to threefold compared to standard Duffing oscillations at similar amplitudes. We further demonstrate, both theoretically and experimentally, that this enhancement is governed by the presence of strong nonlinear damping inherent to these nanoresonators, which controls the phase diffusion of the oscillator.

\begin{figure}[H]
\centering
    \includegraphics[width=\textwidth]{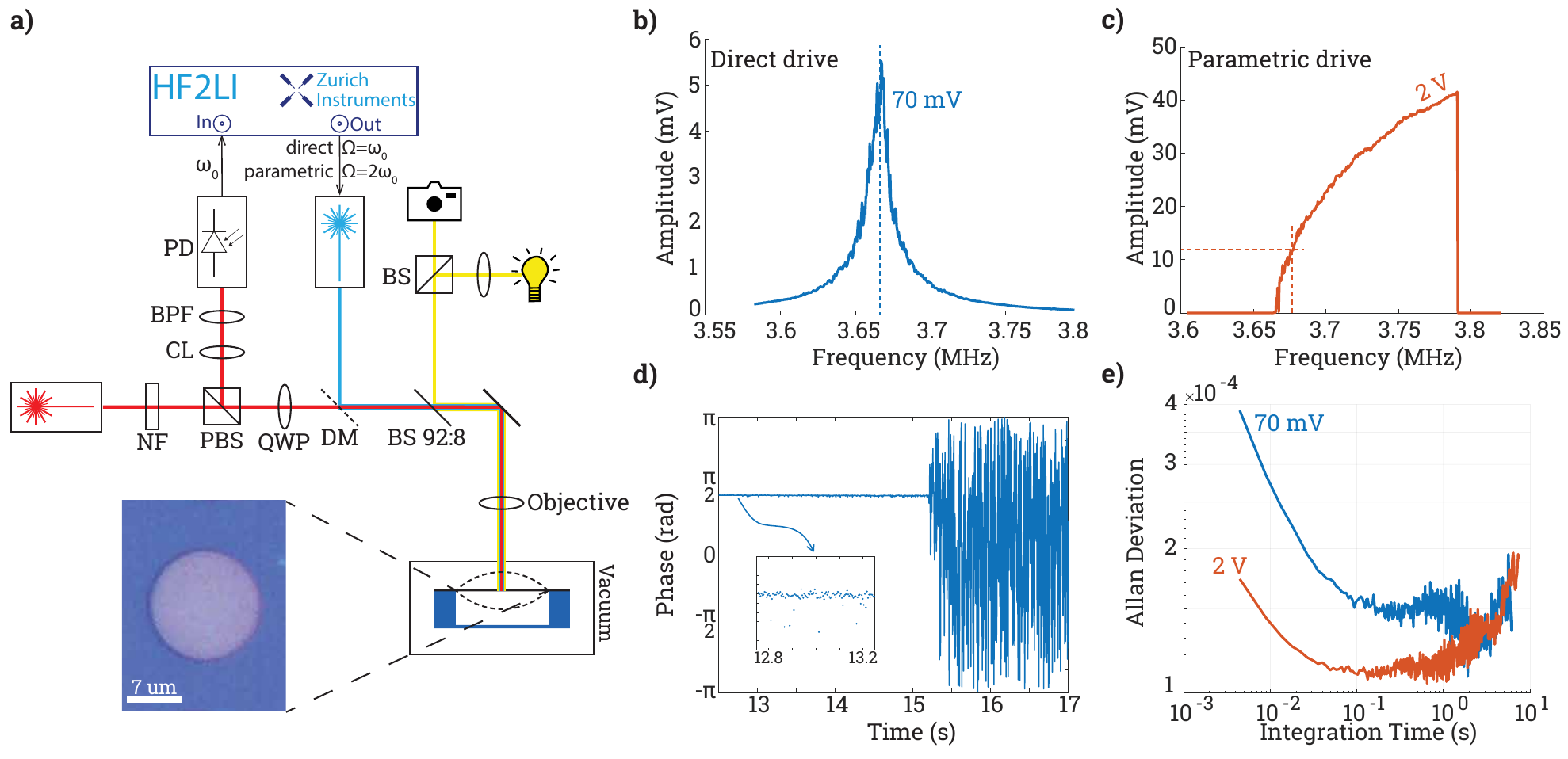}
    \caption{a) Schematic of the experimental setup with a $14\,\mu\mathrm{m}$ diameter bilayer graphene nanodrum placed inside a vacuum chamber (10\textsuperscript{-4} mbar). The optical setup includes a blue laser for optothermal actuation and a red He-Ne laser for the readout. In the schematic, NF stands for neutral filter, PBS polarized beam splitter, CL convergent lens, BPF band-pass filter, PD photodetector, QWP quarter-wave plate, DM dichroic mirror, BS 92:8 pellicle beam splitter, and BS beam splitter. A lock-in amplifier is used for data acquisition and PLL measurements.
b) Directly driven resonance of the graphene nanodrum.
c) Parametric resonance of the graphene nanodrum. 
d) Phase measurement during and after PLL operation. PLL is activated for approximately 15 seconds and then deactivated. The phase-locking is shown in the inset.
e) Allan deviation results of the closed-loop frequency measurements corresponding to the responses in (b) and (c). The PLL operation point for direct-linear response is taken at the peak amplitude (dashed line in (b)), and for the parametric response in the vicinity of the linear resonance frequency (dashed line in (c)).}
    \label{fig:fig1}
\end{figure}

Our experiments are performed on graphene nanodrum resonators fabricated from an ultrathin ($<1\,\mathrm{nm}$) bilayer of chemical vapor--deposited graphene suspended over circular cavities etched in SiO$_2$. The cavities have a diameter of $8\,\mathrm{to}\,14\,\mu\mathrm{m}$ and a depth of $285\,\mathrm{nm}$, forming suspended membranes. To characterize the resonances of these membranes, a customized interferometric setup (\autoref{fig:fig1}a) is used that allows the simultaneous actuation and measurement of the nanodrum's mechanical vibrations \cite{keskekler2021tuning}. To drive the nanodrums both parametrically and directly into resonance, we use a blue diode laser ($\lambda = 488\,\mathrm{nm}$) whose intensity is modulated to provide optothermal tension modulation. To perform the measurements, a red helium–neon laser ($\lambda = 632.8\,\mathrm{nm}$) is focused on the graphene nanodrum, with the back-reflected light collected by a photodetector through a Fabry–Pérot optical cavity. A lock-in amplifier (HF2LI, Zurich Instruments) is further used to measure both the open-loop and closed-loop responses of the nanodrums.

Prior to characterizing the frequency stability of the resonators, their amplitude-frequency response curves are acquired, as shown in \autoref{fig:fig1}b,c. To that end, the signal phase and amplitude registered by the photodetector are recorded while sweeping the modulation frequency of the blue laser. For mechanical characterization, the graphene nanodrum is driven either at the resonator's eigenfrequency ($\Omega \simeq \omega_0$) for direct excitation or at twice the eigenfrequency ($\Omega \simeq 2\omega_0$) for parametric excitation. We note that the direct excitation frequency sweep in \autoref{fig:fig1}b is conducted with a low-amplitude drive and results in a linear response curve that can be fitted with a Lorentzian function. In contrast, the parametric excitation frequency sweep in \autoref{fig:fig1}c is inherently nonlinear as parametric oscillations stem from instability due to a period-doubling bifurcation at $\Omega \simeq 2\omega_0$ and lead to oscillations with frequency $\omega_0$ that are bounded only in the presence of nonlinear damping \cite{lifshitz2008nonlinear}.

After probing the vibrational open-loop response, the phase value for a given frequency is obtained, and then the frequency is tracked over time using a phase-locked loop (PLL) configuration to characterize the frequency stability of the graphene nanodrum in a closed-loop. \autoref{fig:fig1}d demonstrates this resonance frequency tracking upon the activation of the PLL. When activated, the PLL synchronizes the drive frequency to the motion of the nanodrum by maintaining a fixed phase setpoint. This feedback loop continuously adjusts the drive frequency to track the resonance in real-time, creating an active oscillator with self-sustained oscillations. However, when deactivated, the phase evolves more randomly as the drive frequency is not synchronized with the resonator. The PLL-based oscillator is equivalent in terms of frequency stability to other closed-loop techniques, such as self-sustained oscillator (SSO) and feedback-free (FF) schemes, sharing the same steady-state fluctuation performance and speed versus accuracy trade-offs \cite{bevsic2023schemes}. With the frequency readouts as a function of time, the frequency stability is then quantified by computing the Allan deviation $\sigma_y(\tau)$.
\begin{align}
    \sigma_y(\tau) = \sqrt{ \frac{1}{2(N-1)} \sum_{n=1}^{N-1} (y_{n+1} - y_n)^2 },
    \label{eq:allan}
\end{align}
where $y_n$ are the normalized frequency readings averaged over
the $n^{th}$ discrete time intervals of duration $\tau$.

\autoref{fig:fig1}e shows the calculated Allan deviation from closed-loop operations of the direct-driven linear resonance ($V_{\mathrm{drive}} = 70\,\mathrm{mV}$) and of parametric resonance ($V_{\mathrm{drive}} = 2\,\mathrm{V}$). To obtain these, the PLL is operated at $3.67\,\mathrm{MHz}$ for the direct-linear resonance and at twice this frequency for the parametric resonance. The data show that the parametric oscillator exhibits lower Allan deviation, suggesting better frequency stability compared to the direct-driven linear oscillator at fast integration times. It can also be observed that the frequency stability improves with the response amplitude, suggesting that the stability enhancement is driven by the improved signal-to-noise ratio (SNR) resulting from the larger oscillation amplitudes near the onset of the period-doubling bifurcation. The reproducibility of this finding is confirmed across multiple parametrically and directly driven graphene nanodrums operated in closed-loop (see Supporting Information S1 \cite{SI}).

\begin{figure}[H]
\centering
    \includegraphics[width=\textwidth]{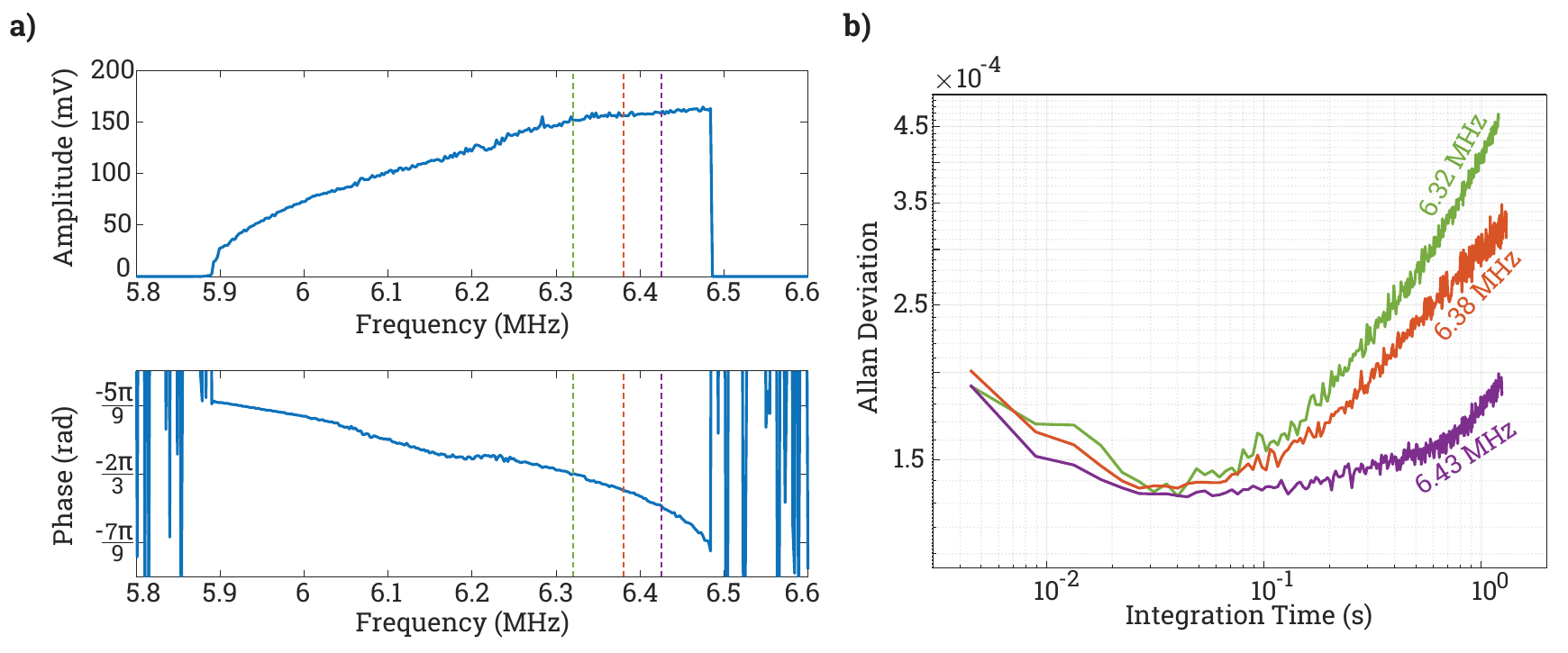}
    \caption{Phase-independent Allan deviation measurements of graphene parametric oscillator. The experiment is performed on a $12\,\mu\mathrm{m}$ diameter bilayer graphene nanodrum.
a) Parametric closed-loop response curve for V\textsubscript{drive} = 2V. Three different PLL operation points are chosen close to the peak amplitude of the response curve (dashed lines), with nearly equal amplitudes but different phases.
b) The Allan deviation results of the closed-loop measurements corresponding to the PLL operation points shown in a).}
    \label{fig:fig2}
\end{figure}

We note that in direct excitation, the feedback drives the resonator through an additive force, resulting in a multiplicative actuation in the phase dynamics. Conversely, in parametric excitation, the feedback drives the resonator through a multiplicative force, resulting in an additive actuation in the phase dynamics. Therefore, we expect that, in contrast to direct excitation, the phase noise (and frequency stability) will not depend on the state (phase) of the feedback loop.
To verify that this is indeed the case, we conducted a set of experiments in which we measured the frequency stability at different operating points along the parametric resonance curve, as shown in \autoref{fig:fig2}a. The acquired amplitude and phase response curves reveal that the phase shift along the upper branch of the parametric resonance is approximately half of the $\pi/2$ shift observed in the direct resonance case.  Note that the phase in \autoref{fig:fig2}a is the directly measured phase. It includes both the resonator response and phase shifts from the experimental setup (such as the external oscillator in the lock-in amplifier), which are assumed to be frequency-independent.
Importantly, the results obtained when computing the Allan deviation in \autoref{fig:fig2}b reveals that different phase values in the feedback loop yield equal minimum Allan deviations at relatively close operation amplitudes, although the long-term stability appears to become better closer to the peak amplitude, possibly because the direct frequency drift generated by slow pump fluctuations is minimized at the peak operation (see Supporting Information S5 for details \cite{SI}).

Our experimental observations in \autoref{fig:fig1} and \autoref{fig:fig2} show that the frequency stability of the parametrically driven oscillator improves with increasing oscillation amplitude and, notably, remains essentially independent of the operational phase. This behavior contrasts with the common notion for nonlinear oscillators, where frequency stability is phase dependent and deteriorates once the oscillator enters the nonlinear regime. Therefore, to gain physical insight into this unconventional trend and to identify the key parameters governing phase diffusion in our system, we introduce the following minimal single-mode theoretical model for our parametrically driven oscillator:
\begin{align}
&\ddot{x}+\omega_0^2(1+\eta(t))x+\gamma x^3= F(t)-2(\Gamma_l+\Gamma_{nl}x^2)\dot{x}+\xi(t),
\label{eq:osc_fast_evo}
\end{align}
where $F(t)=Sx\cos(2\omega_0 t+2\phi+\Delta)$, $x$ is the displacement, and the overdot represents differentiation with respect to time.  The parameters $\Gamma_l$, $\Gamma_{nl}$, $\omega_0$, and $\gamma$ are the linear damping coefficient, nonlinear damping coefficient, natural frequency of the resonator, and Duffing coefficient, respectively. The quantities $S$ and $\Delta$ correspond to the excitation amplitude and phase shift of the drive with respect to the displacement of the resonator, respectively.  Finally, $\eta(t)$ and $\xi(t)$ are the multiplicative (frequency) and additive noise terms, respectively. These noises are assumed to arise from independent physical sources and are, therefore, fully uncorrelated. Specifically, they are assumed to be zero-mean, white, Gaussian, delta-correlated, with intensities denoted by ${\mathcal I}$'s, i.e., $\left< \eta(t) \right> = \left< \xi(t) \right> =\left< \xi(t)\eta(t+\tau) \right>= 0$, $\left< \eta(t)\eta(t+\tau) \right> = 2 {\mathcal I}_{\eta}\delta(\tau)$, and  $\left< \xi(t)\xi(t+\tau) \right> =2 {\mathcal I}_{\xi} \delta(\tau)$.

To analyze the phase dynamics of \autoref{eq:osc_fast_evo}, we use the stochastic averaging method \cite{stratonovich1967topics} (see Supporting Information S2 and S3 \cite{SI}). This asymptotic approach simplifies the dynamics by substituting a slowly varying complex amplitude ansatz into the equation of motion and averaging over the oscillation period to filter out the fast oscillatory terms. The analysis reduces the second-order dynamics into a pair of coupled, first-order Langevin equations with simplified additive noise terms that explicitly describe the slow stochastic evolution of the amplitude ($a$) and phase ($\phi$), which ultimately determine the frequency stability.

Most importantly, the analysis provides estimates of both the Allan deviation and the phase diffusion constant (see Supporting Information S2 \cite{SI}), which are key metrics for determining the frequency stability of the closed-loop oscillator. For convenience, we use the Allan deviation $(\sigma_y)$ to measure frequency stability experimentally, and the diffusion constant $(\rm D_T)$ for the analytical investigation. The value of the diffusion constant quantifies the growth rate of the phase variance, $\langle(\phi -\phi_0)^2 \rangle={\rm D_T} t$, and oscillator linewidth broadening, which measures the degradation in frequency precision due to noise \cite{lax1967classical}. Our analysis gives the following diffusion constant for the parametric oscillator:
\begin{align} 
{\rm D}_{\rm T-Par}(a_{\rm ss})= I_{\phi}(a_{\rm ss})+\left(\frac{3\gamma}{2\omega_0\Gamma_{nl}a_{\rm ss}} \right)^2I_{a}(a_{\rm ss}),
\label{eqn:D_Tp}
\end{align}
where $a_{\rm ss}=\lim_{t\rightarrow\infty}\langle a(t)\rangle$, and $I_{\phi}$ and $I_{a}$ are the amplitude and phase noise intensities from their corresponding Langevin equations, which are given by 
\( I_a(a_{\rm ss})=\omega_0^2 a_{\rm ss}^2{\mathcal I}_\eta/4+{\mathcal I}_\xi/\omega_0^2,\;
I_\phi(a_{\rm ss})=3\omega_0^2{\mathcal I}_\eta/4+{\mathcal I}_\xi/({\omega_0^2 a_{\rm ss}^2}) \) (see Supporting Information S2 for more details \cite{SI}). Inspection of \autoref{eqn:D_Tp} reveals that the diffusion constant of the parametric oscillator depends on the steady-state amplitude ($a_{\rm ss}$) of the resonator but not on the closed-loop operation phase ($\Delta$). This result reinforces the findings from \autoref{fig:fig2}, which showed that short-term parametric frequency stability is phase-independent. Furthermore, the two terms in the expression of the diffusion constant in \autoref{eqn:D_Tp} stem from different origins. The first term, $I_\phi$, results from the projection of the noises $\xi(t)$ and $\eta(t)$ onto phase dynamics, and it is unavoidable in linear $(\gamma=0)$ or nonlinear $(\gamma\neq0)$ oscillators. The second term, $\left[9\gamma^2/(2\omega_0\Gamma_{nl}a_{\rm ss})^2\right]I_{a},$ results from A-F noise conversion \cite{shoshani2024extraordinary,zanette2012frequency,kenig2012optimal} that occurs only in nonlinear oscillators $(\gamma\neq0)$. In fact, the minimalistic model shows that a large nonlinear damping makes the second term in \autoref{eqn:D_Tp} much smaller, thus lowering the total phase diffusion ${\rm D}_{\rm T-Par}$. This behavior of the parametric oscillator is in clear contrast with the diffusion constant of a standard Duffing oscillator, which reads as follows:
\begin{align}
{\rm D}_{\rm T-Duff}=I_\phi(a_{\rm ss})+\left(\frac{3\gamma a_{\rm ss}^3+2S\cos\Delta}{4\omega_0\Gamma_{l} a_{\rm ss}^2}\right)^2I_a(a_{\rm ss}),
\label{eqn:D_Td}
\end{align}
 and which is calculated from Equation \ref{eq:osc_fast_evo} for $F(t)=S\cos(\omega_0 t+\phi+\Delta)$ and $\Gamma_{nl}=0$ using a similar stochastic analysis (see Supporting Information S3 for details \cite{SI}). In particular, we see that unlike the parametric oscillator, here ${\rm D}_{\rm T-Duff}$ depends on the feedback state $(S,\Delta)$, and therefore, the phase shift is usually set to $\Delta=\pi/2$. Moreover, since in a standard Duffing oscillator, the effect of nonlinear damping is negligible (see \cite{palathingal2025axisymmetric}  as well as \autoref{fig:fig4} and its accompanying discussion), we see that for $\Delta=\pi/2$, the two terms in \autoref{eqn:D_Td} represent competing effects. That is, to suppress the additive noise $({\mathcal I}_\xi)$ in the first term, $I_{\phi}$, there is a need to operate with the largest possible amplitude $(a_{\rm ss})$. However, as the amplitude increases, the second term, $\left[3\gamma a_{\rm ss}/(4\omega_0\Gamma_{l} )\right]^2I_a(a_{\rm ss})$, starts to grow due to A-F noise conversion that stems from the oscillator nonlinearity $(\gamma)$. Hence, there is a "sweet spot"  amplitude at which the frequency stability reaches its optimal value, e.g., for ${\mathcal I}_\eta=0$ and ${\mathcal I}_\xi\neq0$, the "sweet spot"  amplitude is given by $a_{\rm ss}^{\rm opt}=2\sqrt{\omega_0\Gamma_l/(3\gamma)}$. This, of course, stands in clear contrast with parametric oscillators, where the presence of nonlinear damping also suppresses the A-F noise conversion, and therefore, in parametric oscillators, higher amplitude translates to better frequency stability with no upper bound or "sweet spot" amplitude. Furthermore, while the present analysis suggests that increasing the oscillation amplitude can further improve the frequency stability of the parametric oscillator, we note that our minimalistic model accounts only for the lowest-order nonlinearities; at larger amplitudes, higher-order nonlinear effects may become important and are not captured by our model.

It is important to clarify the validity range of these theoretical predictions. The derived diffusion coefficient describes frequency stability only for integration times $\tau$ longer than the relaxation time of the oscillator  ($\tau_c=(\Gamma_{nl}a_{\rm ss}^2)^{-1}$) in which small amplitude perturbations $(\delta a)$ decay to zero and the time dependent amplitude $a(t)$ can be replaced with its averaged steady-state value $a_{\rm ss}$. However, at significantly longer timescales, associated with long-term stability, the frequency stability is degraded by deterministic frequency drift mechanisms such as aging and temperature variation, which are not accounted for in this stochastic model. Consequently, the agreement between the theoretical prediction and the experimental data (previously shown in \autoref{fig:fig1}e and \autoref{fig:fig2}b) is strictly valid in the intermediate regime ($\tau_c < \tau < \tau_{\text{drift}}$), where the dynamics is governed by intrinsic phase diffusion. Here, $\tau_{\text{drift}}$ denotes the timescale where long-term drifts dominate over intrinsic phase noise, corresponding to the integration time where the experimental Allan deviation reaches its minimum and drifts with longer gate time.

\begin{figure}[H]
\centering
    \includegraphics[width=\textwidth]{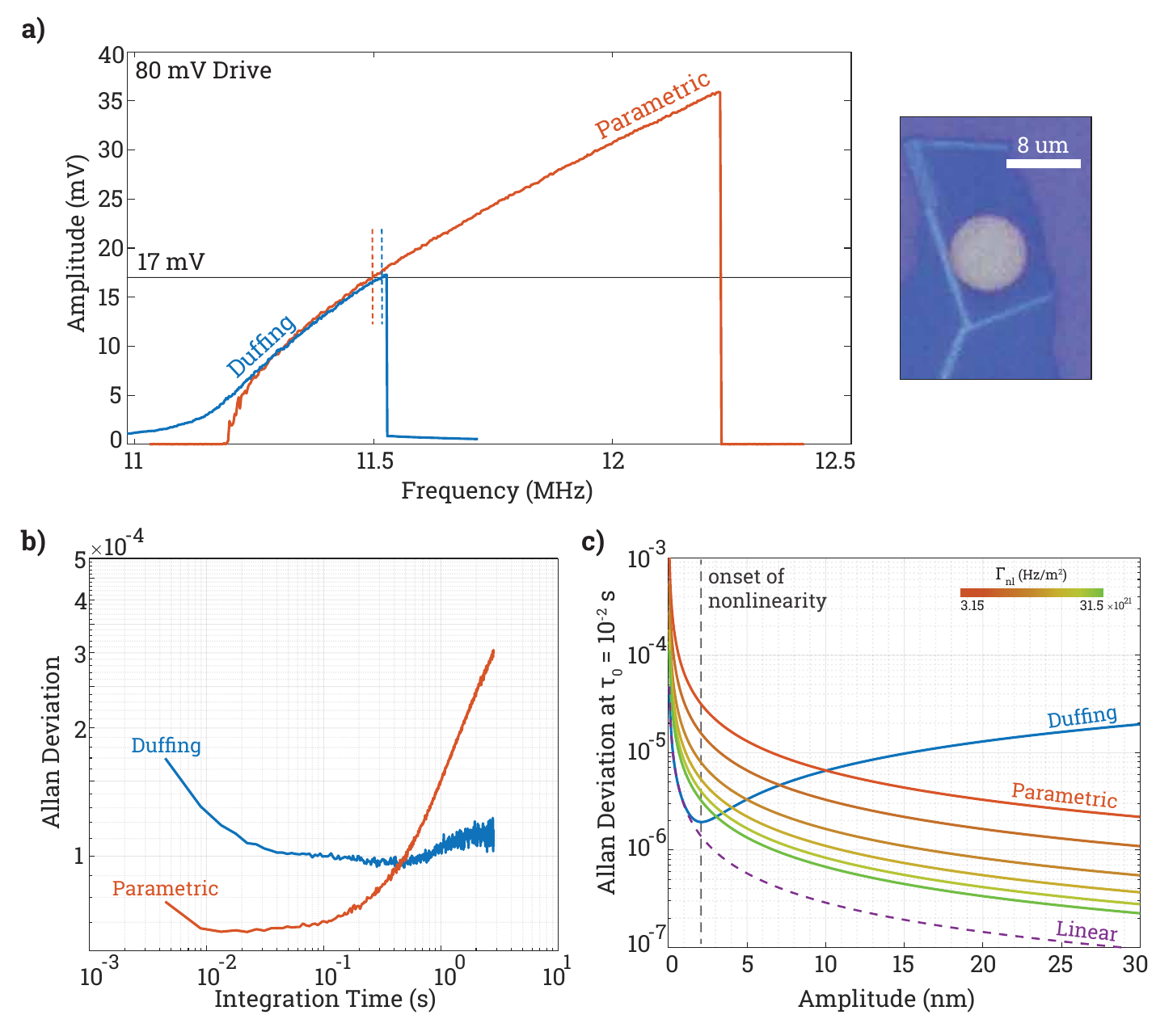}
    \caption{Comparison between the frequency stability of Duffing and parametric oscillators. The experiments are performed for an $8\,\mu\mathrm{m}$ diameter multilayer graphene nanodrum.
a) Frequency sweeps for the Duffing and parametric responses at the same drive level (80 mV). The horizontal line indicates a 17 mV response amplitude, while the color-matching dashed lines mark the PLL operating points. An exfoliated multilayer graphene nanodrum is used to achieve comparable response amplitudes for both Duffing and parametric cases, facilitating a direct comparison at the same drive level.
b) Allan deviation results of the closed-loop frequency measurements corresponding to the responses shown in a).
c) Theoretical frequency stability comparison calculated at a fixed integration time of 10 ms.
The plot compares the stability of parametric and Duffing responses as a function of amplitude, assuming negligible nonlinear damping for the Duffing oscillator. The curves demonstrate that above a certain amplitude threshold, parametric stability is better than the Duffing and follows a similar trend to a linear oscillator, where no A-F noise conversion takes place.}
    \label{fig:fig3}
\end{figure}

\begin{figure}[H]
\centering
    \includegraphics[width=\textwidth]{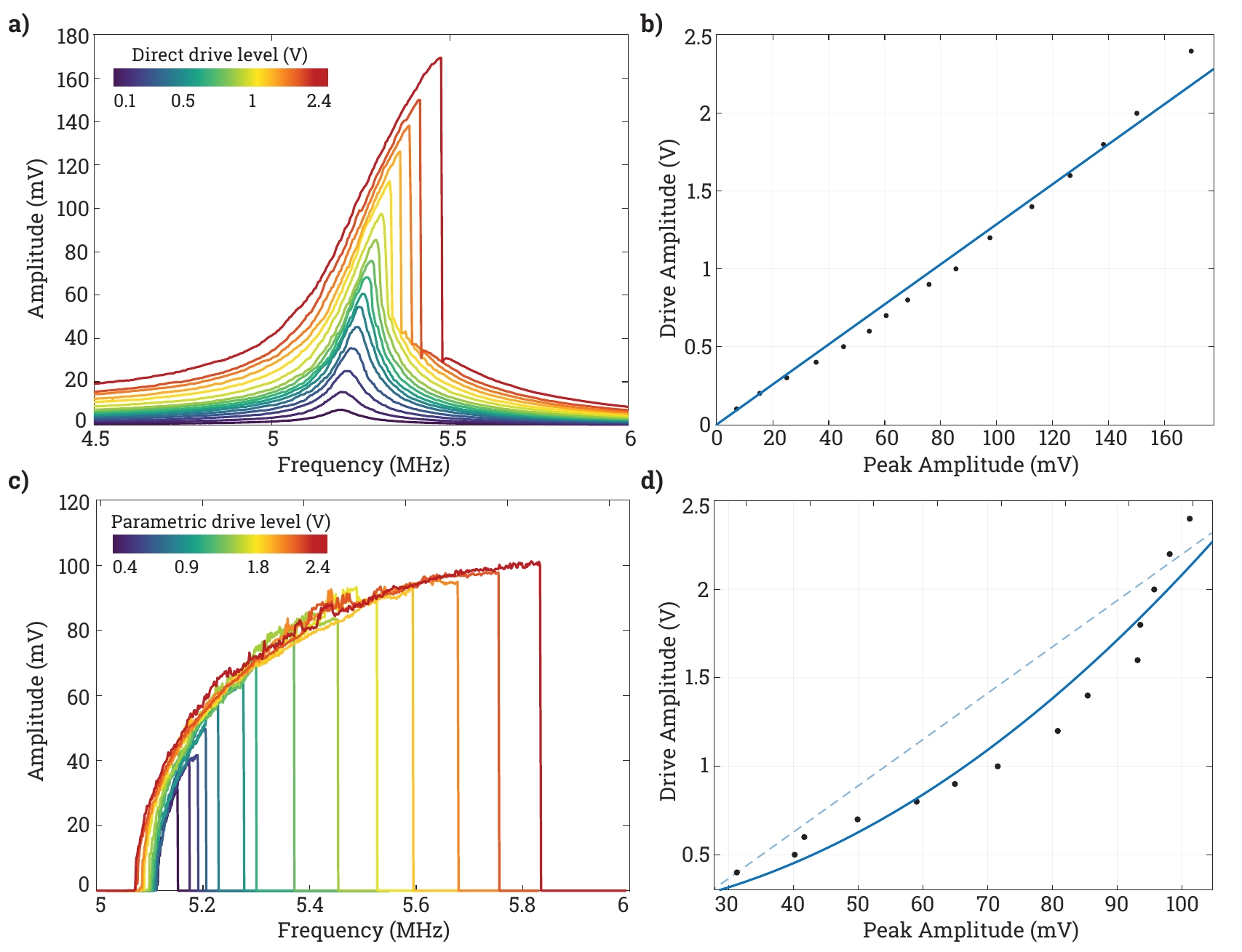}
    \caption{Drive amplitude vs peak amplitude relationships of direct and parametrically driven cases.
a) Direct-driven frequency sweeps with V\textsubscript{drive} from 0.1 V to 2.4 V. These sweeps are taken using a $12\,\mu\mathrm{m}$ diameter bilayer graphene nanodrum.
b) Drive amplitude vs. peak amplitude relationships of the direct-driven responses shown in a). The blue line represents a linear fit ($\mathrm{R^2} \approx 0.99$).
c) Parametrically driven frequency sweeps with V\textsubscript{drive} from 0.4 V to 2.4 V. These sweeps are taken with a $12\,\mu\mathrm{m}$ diameter bilayer graphene nanodrum.
d) Drive amplitude vs. peak amplitude relationships of the parametrically driven responses shown in c). The blue line represents a quadratic fit ($\mathrm{R^2} \approx 0.95$), supporting the model of non-negligible nonlinear damping. The dashed line shows a linear relation, which cannot exist in a bounded parametric response \cite{lifshitz2008nonlinear}, highlighting the deviation from linearity in parametric oscillators.}
    \label{fig:fig4}
\end{figure}

To verify the insights obtained from the theoretical model, we next performed measurements on an exfoliated multilayer graphene nanodrum to compare the frequency stability of Duffing and parametric oscillators. The device was excited using an identical drive voltage of 80 mV under both direct and parametric excitation, and the corresponding frequency response curves were obtained (see \autoref{fig:fig3}a). The phase-locked loop was subsequently set to phase values corresponding to the same response amplitude of 17 mV for both oscillators, allowing a direct comparison based on the diffusion constants defined in \autoref{eqn:D_Tp} and \autoref{eqn:D_Td}. The Allan deviation results in \autoref{fig:fig3}b show that, at identical response amplitudes, the parametric oscillator exhibits a lower Allan deviation compared to the Duffing oscillator for integration times below $\sim 400\,\mathrm{ms}$. This regime is well within the practically relevant operating window for graphene nanodrums, which have a decay time of $\sim 20\,\mu\mathrm{s}$ at room temperature, and therefore confirms the theoretical prediction that parametric excitation can improve short-term frequency stability under comparable conditions. For longer integration times, however, the parametric oscillator displays an earlier onset of frequency drift, leading to degraded long-term stability compared to the directly driven linear and Duffing oscillators (see both \autoref{fig:fig1}e and \autoref{fig:fig3}b).

A possible mechanism of this contrasting short- and long-term behavior can be the slow fluctuations of the actuation parameters that can enter the frequency of the oscillations. As shown in the Supporting Information S5 \cite{SI}, the backbone of the parametrically driven oscillator contains an explicit dependence on the optothermal actuation parameters $(S,\Delta)$. As a result, the slow drift of the actuation amplitude or phase lag generates an additive contribution to the oscillation frequency drift, even when the oscillation amplitude remains unchanged. In contrast, the backbone of a directly driven Duffing oscillator depends only on the oscillation amplitude and does not explicitly depend on $(S,\Delta)$, such that actuator fluctuations influence the oscillation frequency only indirectly through amplitude variations. This difference provides a minimal explanation for the stronger long-term frequency drift observed under parametric excitation. 

To better understand the difference between the short-term stability of Duffing and parametric oscillators, next we performed a simulation based on Equations~\ref{eqn:D_Tp} and \ref{eqn:D_Td}, from which we calculate the short-term Allan deviation at a fixed averaging time $\tau=\tau_0= 10\,\mathrm{ms}$. For simplicity, we confine ourselves to the case of the thermomechanical noise limit and calculate the Allan deviation as below:
\begin{equation} 
\sigma_y^2(\tau)=2\int_0^\infty S_y(f)\frac{\sin^4(\pi f\tau)}{(\pi f\tau)^2}df,
\label{eqn:allan_def}
\end{equation}
where the PSD of fractional-frequency fluctuation $S_y(f)$ is given in Supporting Information S4 \cite{SI} for both parametric and Duffing oscillators. We calculate $\sigma_y(\tau_0)$ using representative graphene nanodrum parameters ($f_0 = 20.13\,\mathrm{MHz}, \gamma = 1.205\times10^{31}\,\mathrm{Hz^2/m^2}, \mathrm{Q = 200}$, as reported by\cite{keskekler2021tuning}), and vary the vibration amplitude and nonlinear damping  $\Gamma_{nl}$ to map the short-term frequency stability. For $\Gamma_{nl}$, particularly, we used the values given in \cite{keskekler2021tuning} up to one order of magnitude higher. The results show that for small vibration amplitudes, the linear regime provides the best stability, whereas beyond the onset of Duffing nonlinearity, the value of the short-term Allan deviation $\sigma_y(\tau_0)$ increases with amplitude due to the A–F effect. In contrast, for the parametric oscillator, the short-term Allan deviation decreases monotonically with increasing amplitude, as nonlinear damping progressively reduces the A–F effect, i.e., the term $\Gamma_{nl}a_{\rm ss}$ suppresses the amplitude sensitivity of the oscillation phase (\autoref{fig:fig3}c). This explains why, at identical response amplitudes, parametric excitation can yield improved short-term frequency stability compared to Duffing operation.

Although \autoref{fig:fig3}b,c demonstrate improved short-term frequency stability of the parametric oscillator, they do not reveal the role of nonlinear damping in setting the short-term Allan deviation. To explicitly probe nonlinear damping in the membrane and assess its impact, we performed additional measurements on the graphene nanodrum as a function of drive amplitude and extracted the evolution of the Duffing and parametric resonances (\autoref{fig:fig4}a,c). The results show that the peak amplitude of the Duffing response varies linearly with the drive strength (\autoref{fig:fig4}b), in excellent agreement with the expected relation $S = 2\omega_0 \Gamma_l a_{\rm ss}^{\rm peak}$, indicating that nonlinear damping is negligible for the drive and response amplitudes used to evaluate the short-term stability in our experiments. In contrast, we notice that the peak amplitude of the parametric resonance exhibits a pronounced quadratic dependence on the drive amplitude (\autoref{fig:fig4}d), consistent with the nonlinear damping backbone relation $S = \omega_0 \Gamma_{nl} (a_{\rm ss}^{\rm peak})^2 + 4\omega_0 \Gamma_l$. As indicated in \autoref{eqn:D_Tp}, this pronounced effect in parametric resonance plays an important role in determining the improved short-term frequency stability of the oscillator. These measurements demonstrate that nonlinear damping manifests much more strongly under parametric excitation, even when direct excitation reaches equal or larger oscillation amplitudes (see also Supporting Information S6 \cite{SI}). 

These findings offer new strategies for enhancing nanomechanical frequency stability, extending it beyond conventional methods that typically operate in the linear regime. While parametric excitation has been explored previously for sensing \cite{papariello2016ultrasensitive}, phase control \cite{kenny2019phase}, phase-noise reduction \cite{villanueva2011nanoscale}, thermomechanical noise-squeezing and signal amplification \cite{grutter1991amplification,suh2010parametric,vinante2013feedback}, the underlying mechanism for closed-loop frequency stability in this regime had remained underexplored. By comparing direct and parametric operation in a closed-loop, this work demonstrates that the improved short-term stability of parametric oscillators is not merely a consequence of signal amplification but is governed by the combination of large displacement amplitudes and nonlinear damping. Crucially, experimental results reveal that nonlinear damping is more pronounced in the parametric regime than in the direct one. Furthermore, our theoretical and experimental analysis confirm that this combination minimizes the A-F noise conversion, enabling the graphene parametric oscillator to suppress phase diffusion more effectively than the direct Duffing oscillator at fast integration times despite higher drift at longer timescales. This capability allows the parametric oscillator to approach the precision limits of linear operation at amplitudes inaccessible to linear resonators. Moreover, this short-term stability enhancement is demonstrated to be robust and independent of the operating phase, effectively turning the intrinsic nonlinearity of graphene into a resource rather than a limitation.
In conclusion, this work highlights the potential of parametric oscillators to surpass the stability limits of conventional operation schemes and paves the way for next-generation 2D material-based oscillators that achieve enhanced precision at fast integration times.\\

\noindent The Supporting Information is available at~\cite{SI}

\begin{quote}
    Additional graphene parametric oscillator PLL measurements, theoretical modeling and analysis of parametric and direct oscillators, Allan variance calculation, discussion on frequency drift due to drive fluctuations, as well as an additional measurement that shows the negligible impact of nonlinear damping in the graphene Duffing oscillator.
\end{quote}

\section{Data availability}
The data that support the findings of this study are available from the corresponding authors upon reasonable request.

\section{Acknowledgment}
Funded/Co-funded by the European Union (ERC Consolidator, NCANTO, 101125458). Views and opinions expressed are, however, those of the author(s) only and do not necessarily reflect those of the European Union or the European Research Council. Neither the European Union nor the granting authority can be held responsible for them. O.S. is grateful for support from the Israel Science Foundation (ISF) grant number 344/22 and the Pearlstone Center of Aeronautical Engineering Studies at BGU. The authors acknowledge SoundCell B.V. for providing the graphene samples for the measurements. We also acknowledge fruitful discussions with Mr. Chris Wattjes and Dr. Tufan Erdogan.

\section{Authors contribution}
{E.K. conducted the measurements, performed experimental data analysis, and fabricated the exfoliated graphene samples; O.S. and E.B. performed theoretical analysis and modeling; A.M.P. built the experimental setup; Data analysis and interpretation were done by E.K., T.M., O.S., and F.A. The project was supervised by T.M. and F.A.; and the manuscript was written by E.K., O.S., T.M., and F.A. with inputs from all authors.}

\section{Competing interests}The authors declare no competing interests.
\newpage
\bibliography{library_used}

@article{villanueva2013surpassing,
  title={Surpassing fundamental limits of oscillators using nonlinear resonators},
  author={Villanueva, LG and Kenig, E and Karabalin, RB and Matheny, MH and Lifshitz, Ron and Cross, MC and Roukes, ML},
  journal={Physical review letters},
  volume={110},
  number={17},
  pages={177208},
  year={2013}
}

@article{kenig2012passive,
  title={Passive phase noise cancellation scheme},
  author={Kenig, Eyal and Cross, MC and Lifshitz, Ron and Karabalin, RB and Villanueva, LG and Matheny, MH and Roukes, ML},
  journal={Physical review letters},
  volume={108},
  number={26},
  pages={264102},
  year={2012},
  publisher={APS}
}

@article{matheny2014phase,
  title={Phase synchronization of two anharmonic nanomechanical oscillators},
  author={Matheny, Matthew H and Grau, Matt and Villanueva, Luis G and Karabalin, Rassul B and Cross, MC and Roukes, Michael L},
  journal={Physical review letters},
  volume={112},
  number={1},
  pages={014101},
  year={2014},
  publisher={APS}
}

@inproceedings{ferre1997draft,
  title={Draft revision of IEEE STD 1139-1988 standard definitions of physical quantities for fundamental, frequency and time metrology-random instabilities},
  author={Ferre-Pikal, ES and Vig, JR and Camparo, JC and Cutler, LS and Maleki, L and Riley, WJ and Stein, SR and Thomas, C and Walls, FL and White, JD},
  booktitle={Proceedings of International Frequency Control Symposium},
  pages={338--357},
  year={1997},
  organization={IEEE}
}

@article{marshall2025high,
  title={High-stability single-ion clock with $5.5\times 10^{-19}$ systematic uncertainty},
  author={Marshall, Mason C and Castillo, Daniel A Rodriguez and Arthur-Dworschack, Willa J and Aeppli, Alexander and Kim, Kyungtae and Lee, Dahyeon and Warfield, William and Hinrichs, Joost and Nardelli, Nicholas V and Fortier, Tara M and others},
  journal={Physical Review Letters},
  volume={135},
  number={3},
  pages={033201},
  year={2025},
  publisher={APS}
}

@article{kenig2012optimal,
  title={Optimal operating points of oscillators using nonlinear resonators},
  author={Kenig, Eyal and Cross, MC and Villanueva, LG and Karabalin, RB and Matheny, MH and Lifshitz, Ron and Roukes, ML},
  journal={Physical Review E},
  volume={86},
  number={5},
  pages={056207},
  year={2012},
  publisher={APS}
}

@article{lax1967classical,
  title={Classical noise. V. Noise in self-sustained oscillators},
  author={Lax, Melvin},
  journal={Physical Review},
  volume={160},
  number={2},
  pages={290},
  year={1967},
  publisher={APS}
}

@book{stratonovich1967topics,
  title={Topics in the theory of random noise},
  author={Stratonovich, Rouslan L},
  volume={2},
  year={1967},
  publisher={CRC Press}
}

@article{bevsic2023schemes,
  title={Schemes for tracking resonance frequency for micro-and nanomechanical resonators},
  author={Be{\v{s}}i{\'c}, Hajrudin and Demir, Alper and Steurer, Johannes and Luhmann, Niklas and Schmid, Silvan},
  journal={Physical Review Applied},
  volume={20},
  number={2},
  pages={024023},
  year={2023},
  publisher={APS}
}

@article{lifshitz2008nonlinear,
  title={Nonlinear dynamics of nanomechanical and micromechanical resonators},
  author={Lifshitz, Ron and Cross, Michael C},
  journal={Reviews of nonlinear dynamics and complexity},
  volume={1},
  number={1},
  year={2008},
  publisher={Wiley Online Library}
}

@article{hinkley2013atomic,
  title={An atomic clock with $10^{-18}$ instability},
  author={Hinkley, Nathan and Sherman, Jeff A and Phillips, Nathaniel B and Schioppo, Macro and Lemke, Nathan D and Beloy, Kyle and Pizzocaro, Marco and Oates, Christopher W and Ludlow, Andrew D},
  journal={Science},
  volume={341},
  number={6151},
  pages={1215--1218},
  year={2013},
  publisher={American Association for the Advancement of Science}
}

@article{vig1999noise,
  title={Noise in microelectromechanical system resonators},
  author={Vig, John R and Kim, Yoonkee},
  journal={IEEE transactions on ultrasonics, ferroelectrics, and frequency control},
  volume={46},
  number={6},
  pages={1558--1565},
  year={1999},
  publisher={IEEE}
}

@misc{li2025cascade,
      title={Cascade of Modal Interactions in Nanomechanical Resonators with Soft Clamping}, 
      author={Zichao Li and Minxing Xu and Richard A. Norte and Alejandro M. Aragón and Peter G. Steeneken and Farbod Alijani},
      year={2025},
      eprint={2507.00805},
      archivePrefix={arXiv},
      primaryClass={cond-mat.mes-hall},
      url={https://arxiv.org/abs/2507.00805}, 
}

@article{grutter1991amplification,
  title = {Mechanical parametric amplification and thermomechanical noise squeezing},
  author = {Rugar, D. and Gr\"utter, P.},
  journal = {Physical Review Letters},
  volume = {67},
  issue = {6},
  pages = {699--702},
  numpages = {0},
  year = {1991},
  month = {Aug},
  publisher = {American Physical Society},
  doi = {10.1103/PhysRevLett.67.699},
  url = {https://link.aps.org/doi/10.1103/PhysRevLett.67.699}
}

@article{keskekler2021tuning,
    author = {Keskekler, A. and Shoshani, O. and Lee, M. and van der Zant and H. S. J. and Steeneken, P. G. and Alijani, F.},
    title = {Tuning nonlinear damping in graphene nanoresonators by parametric–direct internal resonance},
    journal = {Nature Communications},
    year = {2021},
    volume = {12},
    pages = {1099},
    doi = {10.1038/s41467-021-21334-w},
}

@article{sansa2016frequency,
    author = {Sansa, M. and Sage, E. and Bullard, E. C. and Gély, M. and Alava, T. and Colinet, E. and Naik, A. K. and Villanueva, L. G. and Duraffourg, L. and Roukes, M. L. and Jourdan, G. and Hentz, S.},
    title = {Frequency fluctuations in silicon nanoresonators},
    journal = {Nature Nanotechnology},
    year = {2016},
    volume = {11},
    pages = {552--558},
    doi = {10.1038/nnano.2016.19},
}

@article{steeneken2021dynamics,
    author = {Steeneken, P. G. and Dolleman, R. J. and Davidovikj, D. and Alijani F. and van der Zant, H. S. J.},
    title = {Dynamics of 2D material membranes},
    journal =  {2D Materials},
    year = {2021},
    volume = {8},
    pages = {042001},
    doi = {10.1088/2053-1583/ac152c},
}

@article{shoshani2024extraordinary,
    author = {Shoshani, O. and Strachan, S. and Czaplewski, D. and Lopez, D. and Shaw, S. W.},
    title = {Extraordinary frequency stabilization by resonant nonlinear mode coupling},
    journal = {Physical Review Applied},
    volume = {22},
    pages = {054055},
    year = {2024},
    doi = {10.1103/PhysRevApplied.22.054055},
}

@article{zanette2012frequency,
    author       = {Antonio, D. and Zanette, D. H. and López, D.},
    title        = {Frequency stabilization in nonlinear micromechanical oscillators},
    journal      = {Nature Communications},
    volume       = {3},
    pages        = {1--6},
    year         = {2012},
    doi = {10.1038/ncomms1813},
}

@article{kaynak2023paddle,
    author = {Kaynak, B. E. and Alkhaled, M. and Kartal, E. and Yanik, C. and Hanay, M. S.},
    title = {Atmospheric-Pressure Mass Spectrometry by Single-Mode Nanoelectromechanical Systems},
    journal = {Nano Letters},
    volume = {23},
    pages = {8553--8559},
    year = {2023},
    doi = {10.1021/acs.nanolett.3c02343},
}

@article{bachtold2012yoctogram,
    author       = {Chaste, J. and Eichler, A. and Moser, J. and Ceballos, G. and Rurali, R. and Bachtold, A.},
    title        = {A nanomechanical mass sensor with yoctogram resolution},
    journal      = {Nature Nanotechnology},
    volume       = {7},
    pages        = {301--304},
    year         = {2012},
    doi = {10.1038/nnano.2012.42},
}

@article{villanueva2011nanoscale,
    author = {Villanueva, L. G. and Karabalin, R. B. and Matheny, M. H. and Kenig, E. and Cross, M. C. and Roukes, M. L.},
    title = {A Nanoscale Parametric Feedback Oscillator},
    journal = {Nano Letters},
    volume = {11},
    year = {2011},
    doi = {10.1021/nl2031162},
}

@article{kenny2019phase,
  title = {Phase Control of Self-Excited Parametric Resonators},
  author = {Miller, J. M. L. and Shin, D. D. and Kwon, H. and Shaw, S. W. and Kenny, T. W.},
  journal = {Physical Review Appied},
  volume = {12},
  issue = {4},
  pages = {044053},
  year = {2019},
  doi = {10.1103/PhysRevApplied.12.044053},
}

@article{kartal2025reservoir,
    author = {Kartal, E. and Selcuk, Y. and Ahmed, H. and Kaynak, B. E. and Yildiz, M. T. and Erdogan, R. T. and Yanik, C. and Hanay, M. S.},
    title = {Nanomechanical Systems for Reservoir Computing Applications},
    journal = {Advanced Intelligent Systems},
    year = {2025},
    volume = {7},
    issue = {8},
    doi = {10.1002/aisy.202400971},
}

@book{audoin2001time,
    author = {Audoin, C. and Guinot, B.},
    title = {The Measurement of Time: Time, Frequency and the Atomic Clock},
    publisher = {Cambridge University Press},
    year = {2001},
}

@article{dolleman2018opto,
    author = {Dolleman, R. J. and Houri, S. and Chandrashekar, A. and Alijani, F. and van der Zant, H. S. J. and Steeneken, P. G.},
    title = {Opto-thermally excited multimode parametric resonance in graphene membranes},
    journal = {Scientific Reports},
    volume = {8},
    number = {9366},
    year = {2018},
}

@article{ludlow2015optical,
  title = {Optical atomic clocks},
  author = {Ludlow, A. D. and Boyd, M. M. and Ye, J. and Peik, E. and Schmidt, P. O.},
  journal = {Reviews of Modern Physics},
  volume = {87},
  issue = {2},
  pages = {637--701},
  year = {2015},
  publisher = {American Physical Society},
  doi = {10.1103/RevModPhys.87.637},
}

@article{papariello2016ultrasensitive,
  title = {Ultrasensitive hysteretic force sensing with parametric nonlinear oscillators},
  author = {Papariello, Luca and Zilberberg, Oded and Eichler, Alexander and Chitra, R.},
  journal = {Physical Review E},
  volume = {94},
  issue = {2},
  pages = {022201},
  year = {2016},
  publisher = {American Physical Society},
  doi = {10.1103/PhysRevE.94.022201},
}

@article{roslon2022probing,
  title={Probing nanomotion of single bacteria with graphene drums},
  author={Ros{\l}o{\'n}, Irek E and Japaridze, Aleksandre and Steeneken, Peter G and Dekker, Cees and Alijani, Farbod},
  journal={Nature Nanotechnology},
  volume={17},
  number={6},
  pages={637--642},
  year={2022},
  publisher={Nature Publishing Group UK London}
}

@article{vinante2013feedback,
  title = {Feedback-Enhanced Parametric Squeezing of Mechanical Motion},
  author = {Vinante, A. and Falferi, P.},
  journal = {Physical Review Letters},
  volume = {111},
  issue = {20},
  pages = {207203},
  numpages = {5},
  year = {2013},
  month = {Nov},
  publisher = {American Physical Society},
  doi = {10.1103/PhysRevLett.111.207203},
  url = {https://link.aps.org/doi/10.1103/PhysRevLett.111.207203}
}

@article{suh2010parametric,
author = {Suh, Junho and LaHaye, Matthew D. and Echternach, Pierre M. and Schwab, Keith C. and Roukes, Michael L.},
title = {Parametric Amplification and Back-Action Noise Squeezing by a Qubit-Coupled Nanoresonator},
journal = {Nano Letters},
volume = {10},
number = {10},
pages = {3990-3994},
year = {2010},
doi = {10.1021/nl101844r},
    note ={PMID: 20843059},

URL = { 
    
        https://doi.org/10.1021/nl101844r
    
    

},
eprint = { 
    
        https://doi.org/10.1021/nl101844r

}
}

@article{bachtold2022mesoscopic,
  title={Mesoscopic physics of nanomechanical systems},
  author={Bachtold, Adrian and Moser, Joel and Dykman, MI},
  journal={Reviews of Modern Physics},
  volume={94},
  number={4},
  pages={045005},
  year={2022},
  publisher={APS}
}

@article{belardinelli2025hidden,
  title={Hidden Vibrational Bistability Revealed by Intrinsic Fluctuations of a Carbon Nanotube},
  author={Belardinelli, P and Yang, W and Bachtold, A and Dykman, MI and Alijani, Farbod},
  journal={Nano Letters},
  volume={25},
  number={21},
  pages={8443--8449},
  year={2025},
  publisher={ACS Publications}
}

@article{palathingal2025axisymmetric,
  title={Axisymmetric membrane nano-resonators: A comparison of nonlinear reduced-order models},
  author={Palathingal, Safvan and Vella, Dominic},
  journal={International Journal of Non-Linear Mechanics},
  volume={168},
  pages={104933},
  year={2025},
  publisher={Elsevier}
}

@article{arjmandi2025mechanical,
  title={Mechanical Reinforcement of Graphene via Wrinkling},
  author={Arjmandi-Tash, Hadi and Prasad, Roshan and Liu, Hanqing and Verbiest, Gerard and Vella, Dominic and Alijani, Farbod},
  journal={arXiv preprint arXiv:2508.16340},
  year={2025}
}

@misc{SI,
  title={Supporting Information}, 
  author={ },
  OPTyear={2026} 
}


\bigbreak 


\clearpage

\foreach \x in {1,...,7}
{%
\clearpage
\includepdf[pages={\x}]{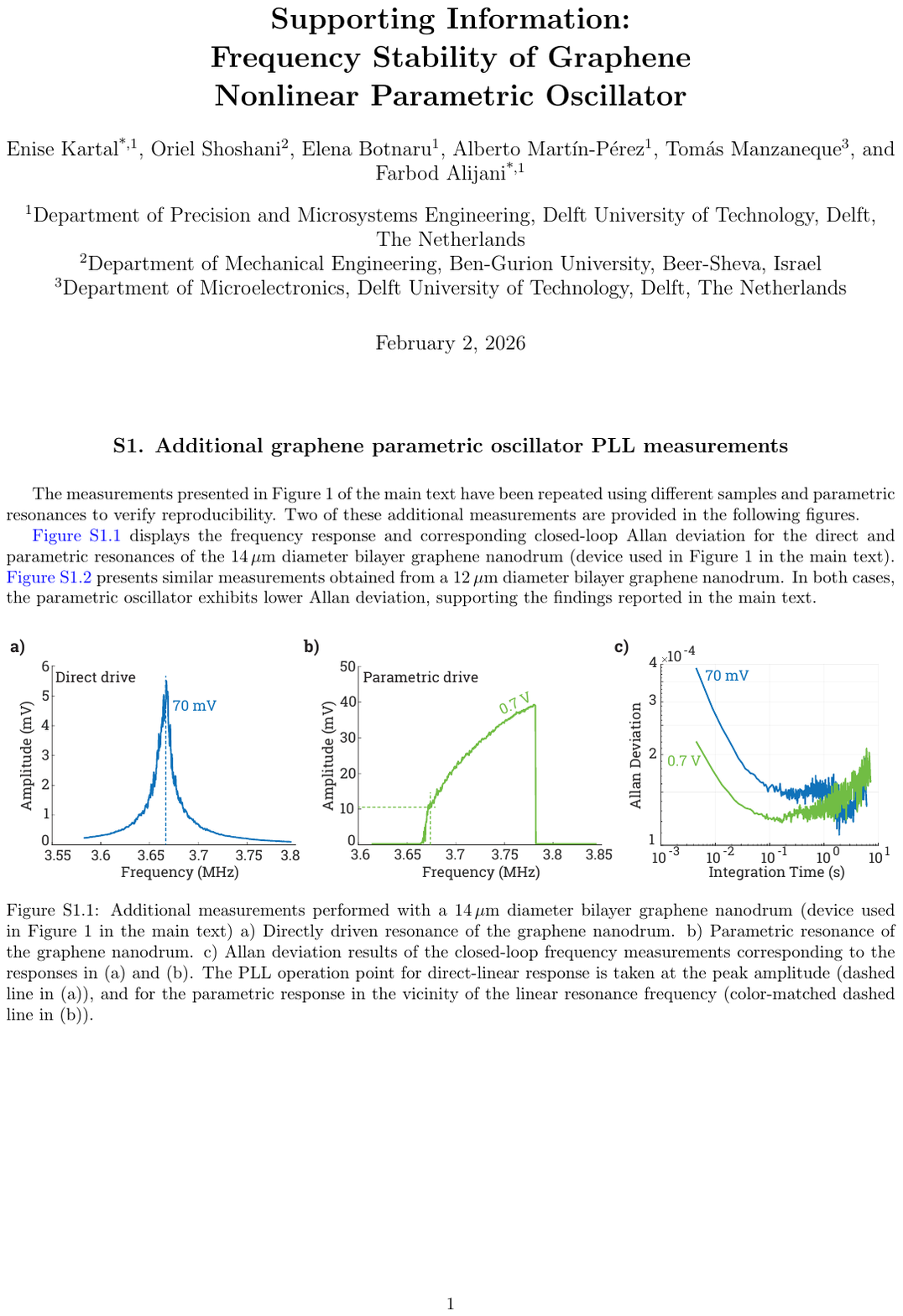} 
}

\end{document}


\label{SI}
\author[*,1]{Enise Kartal}
\author[2]{Oriel Shoshani}
\author[1]{Elena Botnaru}
\author[1]{Alberto Martín-Pérez}
\author[3]{Tomás Manzaneque}
\author[*,1]{Farbod Alijani}

\affil[1]{Department of Precision and Microsystems Engineering, Delft University of Technology, Delft, The Netherlands}
\affil[2]{Department of Mechanical Engineering, Ben-Gurion University, Beer-Sheva, Israel}
\affil[3] {Department of Microelectronics, Delft University of Technology, Delft, The Netherlands}

\maketitle

\renewcommand{\thetable}{S\arabic{section}.\arabic{table}}

\renewcommand{\thefigure}{S\arabic{section}.\arabic{figure}}

\renewcommand{\theequation}{S\arabic{section}.\arabic{equation}}

\renewcommand{\thesection}{S\arabic{section}}

\renewcommand{\refname}{Supplementary References}

\setcounter{section}{1}
\vspace{1em}
\begin{center}
    \large\textbf{S1. Additional graphene parametric oscillator PLL measurements}
\end{center}
\normalsize
\vspace{0.5em}

The measurements presented in Figure 1 of the main text have been repeated using different samples and parametric resonances to verify reproducibility. Two of these additional measurements are provided in the following figures.

\autoref{fig:addt-1} displays the frequency response and corresponding closed-loop Allan deviation for the direct and parametric resonances of the $14\,\mu\mathrm{m}$ diameter bilayer graphene nanodrum (device used in Figure 1 in the main text). \autoref{fig:addt-2} presents similar measurements obtained from a $12\,\mu\mathrm{m}$ diameter bilayer graphene nanodrum.
In both cases, the parametric oscillator exhibits lower Allan deviation, supporting the findings reported in the main text.

\begin{figure}[H]
    \centering
    \includegraphics[width=\linewidth]{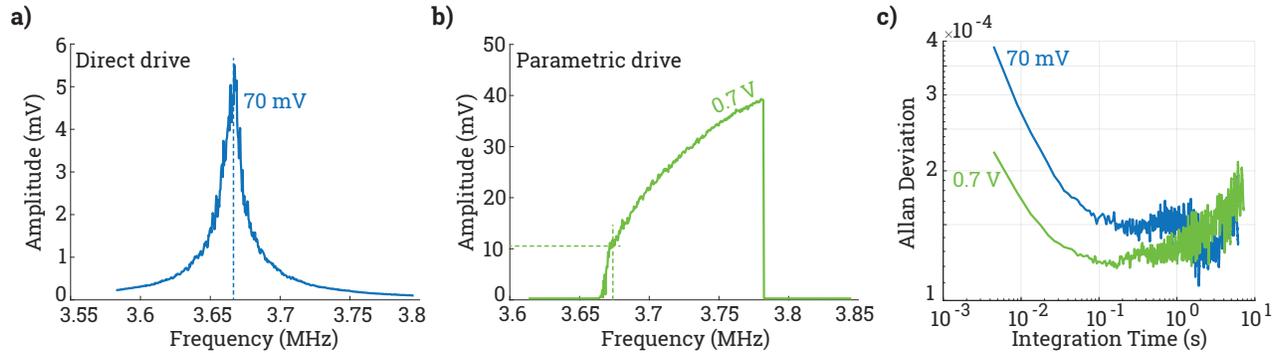}
    \caption{Additional measurements performed with a $14\,\mu\mathrm{m}$ diameter bilayer graphene nanodrum (device used in Figure 1 in the main text)
a) Directly driven resonance of the graphene nanodrum.
b) Parametric resonance of the graphene nanodrum. 
c) Allan deviation results of the closed-loop frequency measurements corresponding to the responses in (a) and (b). The PLL operation point for direct-linear response is taken at the peak amplitude (dashed line in (a)), and for the parametric response in the vicinity of the linear resonance frequency (color-matched dashed line in (b)).}
    \label{fig:addt-1}
\end{figure}

\begin{figure}[H]
    \centering
    \includegraphics[width=\linewidth]{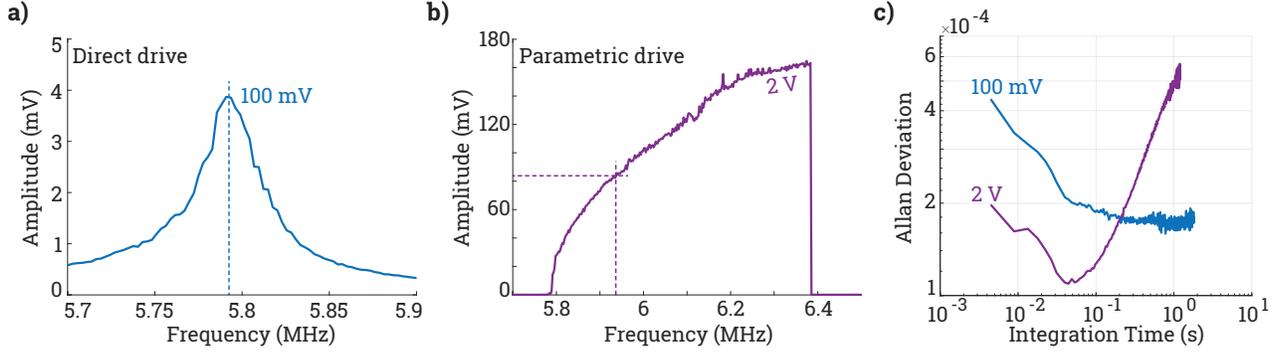}
    \caption{Additional measurements performed with a $12\,\mu\mathrm{m}$ diameter bilayer graphene nanodrum
a) Directly driven resonance of the graphene nanodrum.
b) Parametric resonance of the graphene nanodrum. 
c) Allan deviation results of the closed-loop frequency measurements corresponding to the responses in (a) and (b). The PLL operation point for direct-linear response is taken at the peak amplitude (dashed line in (a)), and for the parametric response in the vicinity of the linear resonance frequency (color-matched dashed line in (b)).}
    \label{fig:addt-2}
\end{figure}

\setcounter{section}{2}
\vspace{1em}
\begin{center}
    \large\textbf{S2. Minimalistic stochastic model of a parametric oscillator}
\end{center}
\normalsize
\vspace{0.5em}

\setcounter{subsection}{0}
\setcounter{equation}{0}
\setcounter{figure}{0}
\setcounter{table}{0}

As described in the main text, we consider the following mathematical model to describe the parametric oscillator dynamics. 
\begin{align}
&\ddot{x}+\omega_0^2(1+\eta(t))x+\gamma x^3= Sx\cos(2\omega_0 t+2\phi+\Delta)-2(\Gamma_l+\Gamma_{nl}x^2)\dot{x}+\xi(t).
\label{eq:osc_fast_evo}
\end{align}

We note that while the noises are taken to be ideal (i.e., zero-mean, white, Gaussian, delta-correlated, and independent), the idealization of the noises can be used for general stationary zero-mean Non-Gaussian noises, as long as their correlation times are considerably shorter than the relaxation time of the resonator, e.g.,  $\tau_{\rm cor}=(1/\langle \xi^2\rangle)\int_0^\infty\left< \xi(t))\xi(t+\tau) \right>d\tau\ll1/(\Gamma_{nl}a_{\rm ss}^2)$. \\

For weak nonlinearity ($|\gamma x^2| \ll \omega_0^2,~|\Gamma_{nl}x^2|\ll\omega_0 $), weak drive ($S\ll\omega_0^2$), slow linear decay ($\Gamma_l \ll \omega_0$), and small noise (defined in the sequel), we have a weakly nonlinear oscillator, and therefore, its amplitude and phase vary on a slow timescale. Thus, we make the ansatz $x(t)= A(t) \ee^{\ii \omega_0 t}+A^*(t) \ee^{-\ii \omega_0 t}$, and $\dot x(t)=  \ii \omega \left[ A(t) \ee^{\ii \omega_0 t}-A^*(t) \ee^{-\ii \omega_0 t} \right]$, where the slowly varying complex amplitude $A(t)$ and its complex conjugate $A^*(t)$, satisfy the constraint $\dot A(t) \ee^{\ii \omega_0 t}+ \dot A^*(t) \ee^{-\ii \omega_0 t}=0$. By using this constraint to eliminate $\dot A^*$, and neglecting fast oscillatory terms, we obtain the following complex-amplitude equation 
\begin{align}
&\dot A = -\left[\Gamma_l+\left(\Gamma_{nl}-\frac{3\ii\gamma}{2\omega_0}\right)|A|^2\right] A-\frac{\ii S A^*}{4 \omega_0}\ee^{\ii (\Delta +\phi)} -\frac{\ii }{2 \omega_0}\langle\xi \ee^{-\ii\omega_0 t}\rangle_{\rm Tp}+ \frac{\ii \omega_0}{2}\langle \eta(A + A^*\ee^{-2\ii\omega_0 t})\rangle_{\rm T_p}, 
\label{eqn:Adot}  
\end{align}
where $\phi(t)=\angle{A(t)}$, and $\langle\{\cdot\}\rangle_{\rm T_p}$ denotes the time-averaged value of $\{\cdot\}$ over the period ${\rm T_p}=2\pi/\omega_0$. \\

Using the polar notation, $A(t)=a(t) \ee^{\ii \phi(t)}/2$, we obtain the following slowly varying amplitude and phase equations
\begin{align}
\dot a =&- \left(\Gamma_l+\frac{\Gamma_{nl}}{4}a^2-\frac{S\sin\Delta}{4\omega_0} \right)a +\frac{3\omega_0^2{\mathcal I}_\eta}{8}a+\frac{{\mathcal I}_\xi}{2\omega_0^2 a}+\sqrt{I_a(a)}\chi_1(t),   \label{eqn:adot} \\
\dot \phi =& \frac{1}{4\omega_0 } \left(\frac{3\gamma a^2}{2} -S \cos\Delta \right)+\sqrt{I_\phi(a)}\chi_2(t) \label{eqn:phidot}
\end{align}

in which \( I_a(a_{\rm ss})=\omega_0^2 a_{\rm ss}^2{\mathcal I}_\eta/4+{\mathcal I}_\xi/\omega_0^2,\;
I_\phi(a_{\rm ss})=3\omega_0^2{\mathcal I}_\eta/4+{\mathcal I}_\xi/({\omega_0^2 a_{\rm ss}^2}) \), and where $a_{\rm ss}=\lim_{t\rightarrow\infty}\langle a(t)\rangle$. The simplified noise terms $\chi_{1,2}(t)$ in Eqs. \eqref{eqn:adot}-\eqref{eqn:phidot}, which have the properties: $\langle\chi_{1,2}(t)\rangle=0$ and $\langle\chi_{k}(t)\chi_{l}(t+\tau)\rangle=\delta_{kl}\delta(\tau)$,  are obtained from the method of stochastic averaging using the fact that the correlation times of $\xi(t)$ and $\eta(t)$ are significantly smaller than the relaxation time of the oscillator, that is, they are effectively delta-correlated. 

\subsection{Deterministic response}

The noise-free $({\mathcal I}_\xi={\mathcal I}_\eta=0)$ steady operating amplitude $a(t)=a_{\rm ss}\neq0$ represents the state about which fluctuations occur.  It is found by solving for the constant amplitude $ a_{\rm ss}$ when setting $\dot a$ in Eq. (\ref{eqn:adot}) to zero, which yields 
\begin{align}
    a_{\rm ss}=\sqrt{\frac{S}{\omega_0\Gamma_{nl}}\left(\sin \Delta-\frac{4\omega_0\Gamma_l}{S}\right)}.
    \label{eq:a_ss}
\end{align}
 
 Note that the non-trivial steady operating amplitude $a(t)=a_{\rm ss}\neq0$ exists only when $S>S_{\rm cr}=4\omega_0\Gamma_l$ (and $\Gamma_{nl}\neq0$), when $S\leq S_{\rm cr}$, there is only a trivial response $a=0$.   The noise-free oscillator frequency, denoted as $\Omega$, is given by adding $\omega_0$ to the steady value of $\dot \phi$, which, using $a(t)=a_{\rm ss}$  in Eq. (\ref{eqn:phidot}), is found to be
$$\Omega=\omega_0+\frac{1}{4 \omega_0 } \left(\frac{3\gamma S\sin \Delta}{2\omega_0\Gamma_{nl}}-\frac{6\gamma\Gamma_l}{\Gamma_{nl}}-S \cos\Delta\right).$$

An alternative way to express the steady operating conditions, one that is useful for comparison with other situations, is obtained from the deterministic part of Eqs. (\ref{eqn:adot})-(\ref{eqn:phidot}), by uncoupling the phase variable $\Delta$, yielding
 \begin{align}
    \left(\frac{S}{4\omega_0}\right)^2&=\left[\left(\Omega-\omega_0- \frac{3\gamma a_{\rm ss}^2}{8\omega_0}\right)^2+
    \left(\Gamma_l+\frac{\Gamma_{nl}a_{\rm ss}^2}{4}\right)^2\right],\label{eqn:res_curve}\\
    \tan\Delta&=-\left(\Gamma_l+\frac{\Gamma_{nl}a_{\rm ss}^2}{4}\right)\left(\Omega-\omega_0- \frac{3\gamma a_{\rm ss}^2}{8\omega_0}\right)^{-1}. \label{eqn:phase_res}
 \end{align}
 
\begin{figure}[H]
    \centering
    \includegraphics[width=0.3\linewidth]{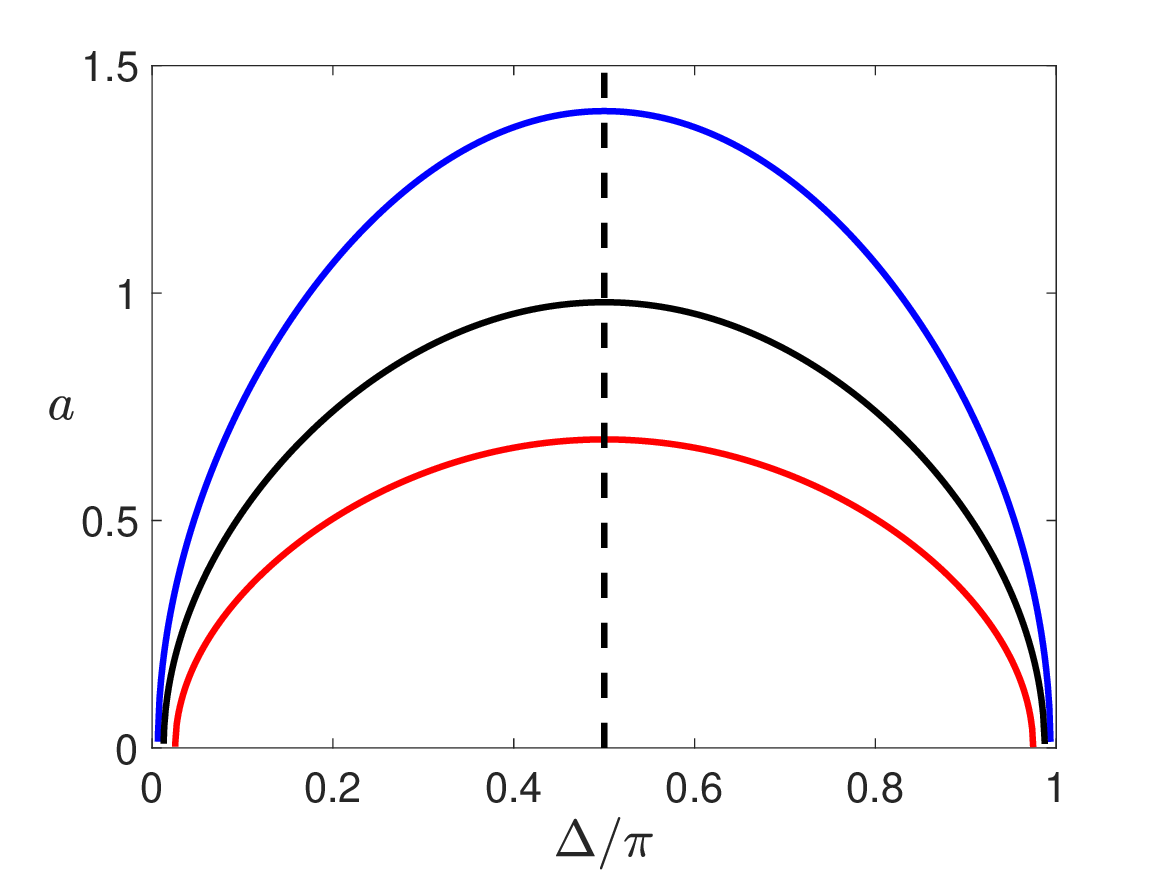}
     \includegraphics[width=0.3\linewidth]{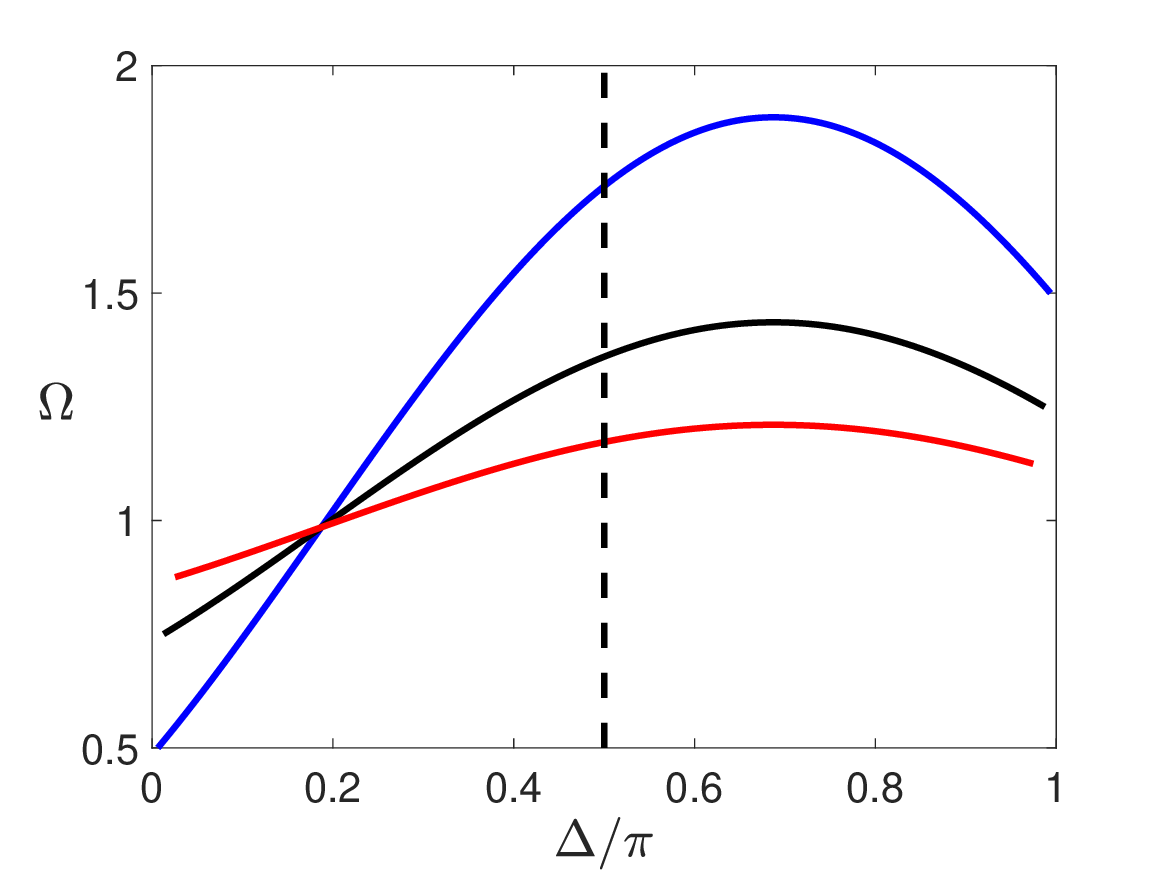}
       \includegraphics[width=0.3\linewidth]{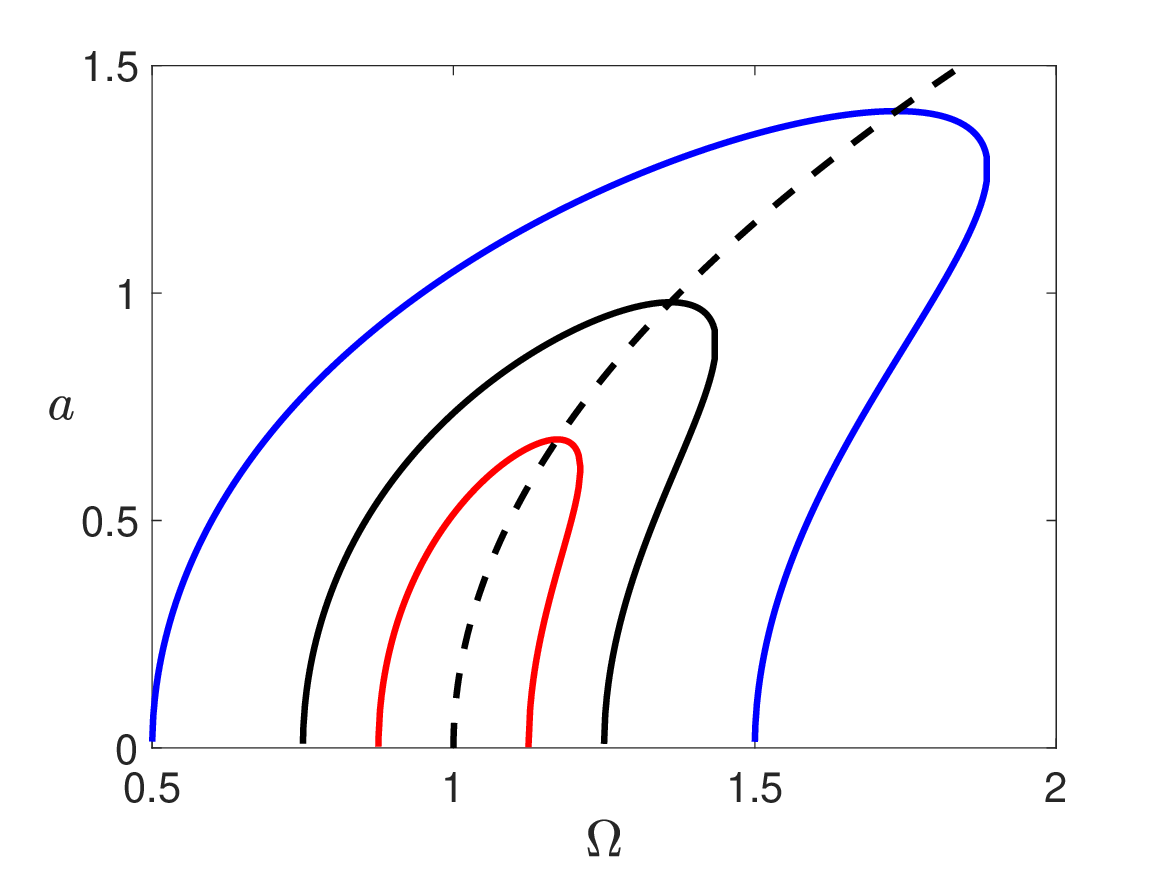}
    \caption{The deterministic response curves for different drive amplitudes [$S=0.5$ (red), $S=1$ (black), $S=2$ (blue)], with fixed system parameters [$\omega_0=1, \gamma=1, \Gamma_l=0.01, \Gamma_{nl}=1$]. As the phase shift $\Delta$ varies from 0 to $\pi$, the amplitude (left panel) and frequency (center panel) of the oscillator are uniquely determined, and the amplitude-frequency response curve (left panel) is a  single-valued function of the phase $\Delta$ for a given $S$.}
    \label{fig:res_curves}
\end{figure}

Eqs. \eqref{eqn:res_curve}-\eqref{eqn:phase_res} are identical to the steady-state conditions obtained for a parametrically driven (open-loop) nonlinear resonator with $\Omega$ and $S$ playing the roles of the drive frequency and amplitude, respectively. Therefore, the response curves of the closed-loop oscillator are identical in shape to the response curves of a harmonically driven open-loop resonator. However, a significant difference between open-loop and closed-loop operations is that in an open-loop system, the drive frequency $\Omega$ and amplitude $S$ are fixed and the amplitude $a_{\rm ss}$ and phase $\Delta$ are determined from the steady-state conditions, yielding response curves with multiple co-existing non-trivial stable and unstable branches.  Conversely, in the closed-loop system, we fix the drive phase $\Delta$ and amplitude $S$ and calculate the frequency $\Omega$ and amplitude $a_{\rm ss}$. Hence, for a closed-loop system, both non-trivial amplitude and frequency are single-valued functions of the phase $\Delta$ for a given $S$, as shown in Fig. \ref{fig:res_curves}. 


\subsection{Small noise analysis of parametric oscillator}

At this point, we can quantify what is meant by small noise intensity, namely, ${\mathcal I}_{\eta}\ll (\Gamma_{nl}a_{\rm ss}^2/\omega_0)^{2}$ for the frequency noise and ${\mathcal I}_\xi\ll(\Gamma_{nl}\omega_0 a_{\rm ss}^3)^2$ for the thermal noise (i.e., $|I_a(a_{\rm ss})|\ll|\Gamma_{nl}a_{\rm ss}^3|^2$). For stability and noise considerations, we introduce deviations from the noise-free steady operation state as $a(t)=\delta a(t) +a_{\rm ss}, ~\dot\phi(t)=\delta\dot\phi(t)+\Omega$, and substitute these into Eqs. (\ref{eqn:adot})-(\ref{eqn:phidot}), assume small deviations, linearize in the deviation terms, and neglect the noise terms $3\omega_0^2{\mathcal I}_\eta a/8+{\mathcal I}_\xi/(2\omega_0 a)$ in Eq. \eqref{eqn:adot}, as these are much smaller than the nonlinear damping of the oscillator, which sets the value of steady-state amplitude $a_{\rm ss}$.  This yields the following pair of linear Langevin equations describing the stochastic dynamics local to the noise-free steady operating condition,
\begin{align}
\delta\dot a =&- \frac{\Gamma_{nl}a_{\rm ss}^2}{2}\delta a+I_{a}^{1/2}(a_{\rm ss})\chi_1(t),   \label{eqn:delta_adot} \\
  \delta\dot\phi =& \frac{3\gamma a_{\rm ss}}{4 \omega_0 } \delta a+I_{\phi}^{1/2}(a_{\rm ss})\chi_2(t). \label{eqn:delta_phidot}
  \end{align}

The amplitude equation is uncoupled from the phase equation. Thus, Eq. (\ref{eqn:delta_adot}) can be readily integrated to yield
 \begin{align}
\delta a(t)&=\delta a(0)e^{-\Gamma_{nl}a_{\rm ss}^2 t/2}+ I_{a}^{1/2}( a_{\rm ss})\int_0^t\chi_1(\tau)e^{\Gamma_{nl}a_{\rm ss}^2(\tau-t)/2}d\tau,
\label{eq:delta a(t)}
\end{align}
with a mean value $\langle\delta a(t)\rangle=\delta a(0)e^{-\Gamma_{nl}a_{\rm ss}^2 t/2}$.
Hence, by substitution of Eq. (\ref{eq:delta a(t)}) into Eq. (\ref{eqn:delta_phidot}) and integration with respect to time, we find 
 \begin{align}
&\delta \phi(t)=\langle\delta\phi(t)\rangle+ I_{\phi}^{1/2}( a_{\rm ss})\int_0^t\chi_2(\tau)d\tau+\frac{3\gamma a_{\rm ss}I_{a}^{1/2}(a_{\rm ss})}{4\omega_0 } \!\int_0^t\!\int_0^\tau\!\chi_1(\rho)e^{\Gamma_{nl}a_{\rm ss}^2(\rho-\tau)/2}d\rho d\tau,
\label{eq:delta phi(t)}
\end{align}
where 
\begin{align}
  \langle\delta\phi(t)\rangle&=\delta\phi(0)+ \frac{3\gamma \delta a(0) }{ 2\omega_0 \Gamma_{nl}a_{\rm ss}}(1-e^{-\Gamma_{nl}a_{\rm ss}^2 t/2}).
\end{align}

Thus, the variances of $\delta a$ and $\delta\phi$ are given by
\begin{align}
\langle \delta a^2(t)\rangle&-\langle \delta a(t)\rangle^2=\frac{2I_{a}(a_{\rm ss})}{2\Gamma_{nl}a_{\rm ss}^2}(1-e^{-\Gamma_{nl}a_{\rm ss}^2 t}),
\label{eq:a var}\\
\langle \delta \phi^2(t)\rangle&-\langle \delta \phi(t)\rangle^2=\left[I_{\phi}(a_{\rm ss})+\left(\frac{3\gamma}{2\omega_0\Gamma_{nl}a_{\rm ss}} \right)^2I_{a}(a_{\rm ss})\right]t.
\label{eq:phi var}
\end{align} 

\emph{\textbf{Phase diffusion:}} From Eqs. \eqref{eq:a var}-\eqref{eq:phi var} we see that  while the steady-state variance of the amplitude remains constant and small, $\lim\limits_{t \to \infty}\langle \delta a^2\rangle=2I_{a}/(\Gamma_{nl}a_{\rm ss}^2)$, the phase is unconstrained and running freely with a variance that increases linearly in time, $\langle (\phi-\phi_0)^2\rangle=D_{\rm T}(a_{\rm ss})t$, where 
\begin{equation} 
{\rm D_{T-Par}}(a_{\rm ss})= I_{\phi}(a_{\rm ss})+\left(\frac{3\gamma}{2\omega_0\Gamma_{nl}a_{\rm ss}} \right)^2I_{a}(a_{\rm ss})
\label{eqn:D_Tp}
\end{equation}
is the total diffusion constant of the phase.\\

\emph{\textbf{Noise induced spectral line broadening:}} Unlike an open-loop driven resonator, where the dissipation in the resonator determines the linewidth of the spectral response, the linewidth of a closed-loop oscillator is nonzero only in the presence of noise. That is, the spectral line is broadened only if there is some stochastic influence that leads to frequency fluctuations. Therefore, the power spectral density of the oscillator $S_x(f)$ can also be used to estimate frequency stability. In particular, if we neglect the amplitude fluctuations, which are small in comparison to the phase fluctuations, we get an exponentially decaying autocorrelation function for the output signal of the oscillator. 
\begin{equation}
R(\tau)=\langle x(t)x(t+\tau)\rangle=\frac{1}{2}a_{\rm ss}^{2}\ee^{-D_{\rm T}|\tau|/2}\cos(\Omega\tau),
\label{eq:autocorr}
\end{equation}
which corresponds to the following power spectral density
\begin{align}
&S_x(f)=2\int_0^{\infty}\left\langle x(t)x(t+\tau) \right\rangle\cos(2\pi f\tau)d\tau=a_{\rm ss}^{2}D_{\rm T}\frac{(2\pi f)^2+\Omega^{2}+(D_{\rm T}/2)^2}{[(2\pi f)^2-\Omega^{2}-(D_{\rm T}/2)^2]^{2}+(2\pi f)^2D_{\rm T}^{2}}.
\label{eq:PSD}
\end{align}

Since $D_{\rm T}\ll\Omega$, the spectral-density is concentrated in a narrow frequency band $|\omega-\Omega|\sim D_{\rm T}$, and reduces to the well-know Lorentzian distribution,
\begin{equation}
S_x(f)\approx a_{\rm ss}^{2}\frac{D_{\rm T}/2}{(2\pi f-\Omega)^{2}+(D_{\rm T}/2)^2}.
\label{eq:reduced PSD}
\end{equation}

The total power of the signal $P=\int_{0}^{\infty}S_x( f)df= a_{\rm ss}^2/2$ is unaltered by the presence of the phase noise. However, in contrast to the purely harmonic signal, for which $D_{\rm T}\rightarrow0$ corresponding to a delta-function spectral-density centered at $2\pi f=\Omega$, in the noisy signal, with $D_{\rm T}\neq0$, the spectral line is broadened by phase diffusion with full line width of $D_{\rm T}$ at half maximum. 

\setcounter{section}{3}
\setcounter{subsection}{0}
\setcounter{equation}{0}
\setcounter{figure}{0}
\setcounter{table}{0}
\vspace{1em}
\begin{center}
    \large\textbf{S3. Noise analysis of the direct Duffing oscillator}
\end{center}
\normalsize
\vspace{0.5em}

To analyze the frequency stability of the direct Duffing oscillator, we consider the following mathematical model
\begin{align}
&\ddot{x}+\omega_0^2(1+\eta(t))x+\gamma x^3= S\cos(\omega_0 t+\phi+\Delta)-2\Gamma_l\dot{x}+\xi(t).
\label{eq:osc_fast_evo}
\end{align}

Using a similar analysis to the case of a parametric oscillator, we obtain the following pair of Langevin equations 
\begin{align}
\delta\dot a =&- \Gamma_l\delta a+I_{a}^{1/2}(\bar a)\chi_1(t),   \label{eqn:delta_adot_duff} \\
\delta\dot\phi =& \frac{1}{2 \omega_0 } \left( \frac{3\gamma a_{\rm ss}}{4}+ \frac{S \cos\Delta}{a_{\rm ss}^2} \right)\delta a+I_{\phi}^{1/2}(\bar a)\chi_2(t). \label{eqn:delta_phidot_duff}
  \end{align}

By integrating Eqs. \eqref{eqn:delta_adot_duff}-\eqref{eqn:delta_phidot_duff}, we find that 
\begin{align}
\delta a(t)&=\delta a(0)e^{-\Gamma_l t}+ I_{a}^{1/2}(a_{\rm ss})\int_0^t\chi_1(\tau)e^{\Gamma_l(\tau-t)}d\tau,
\label{eq:delta a(t) duff}\\
\delta \phi(t)&=\langle\delta\phi(t)\rangle+ I_{\phi}^{1/2}(a_{\rm ss})\int_0^t\chi_2(\tau)d\tau\nonumber\\
&\!+\frac{I_{a}^{1/2}(a_{\rm ss})}{2 \omega_0 } \left( \frac{3\gamma a_{\rm ss}}{4}+ \frac{S \cos\Delta}{a_{\rm ss}^2} \right)\!\int_0^t\!\int_0^\tau\!\chi_1(\rho)e^{\Gamma_l(\rho-\tau)}d\rho d\tau,
\label{eq:delta phi(t)}
\end{align}
where 
\begin{align}
  \langle\delta\phi(t)\rangle&=\delta\phi(0)+ \frac{\delta a(0) }{2 \omega_0 \Gamma} \left(\frac{3\gamma a_{\rm ss}}{4}+ \frac{S \cos\Delta}{a_{\rm ss}^2} \right)(1-e^{-\Gamma_l t}).
\end{align}

Hence, the diffusion constant of a Duffing oscillator is given by
\begin{align}
{\rm D}_{\rm T-Duff}=I_\phi(a_{\rm ss})+\left(\frac{3\gamma a_{\rm ss}^3+2S\cos\Delta}{4\omega_0\Gamma_l a_{\rm ss}^2}\right)^2I_a(a_{\rm ss}).  
\label{eqn:D_Td}
\end{align}

By setting $\Delta=\pi/2$, we find that 
\begin{align}
\frac{{\rm D}_{\rm T-Par}-I_\phi(a_{\rm ss})}{{\rm D}_{\rm T-Duff}-I_\phi(a_{\rm ss})}=\left(\frac{2\Gamma_l}{\Gamma_{nl}a_{\rm ss}^2}\right)^2.   
\label{eqn:D_Tratio}
\end{align}

\setcounter{section}{4}
\setcounter{subsection}{0}
\setcounter{equation}{0}
\setcounter{figure}{0}
\setcounter{table}{0}
\vspace{1em}
\begin{center}
    \large\textbf{S4. Allan variance calculation}
\end{center}
\normalsize
\vspace{0.5em}

\subsection{Allan variance of parametric oscillator}

To calculate the Allan deviation, we use its definition  
\begin{equation} 
\sigma_y^2(\tau)=2\int_0^\infty S_y(f)\frac{\sin^4(\pi f\tau)}{(\pi f\tau)^2}df,
\label{eqn:allan_def}
\end{equation}
where $S_y(f)$ is the power spectral density of fractional-frequency fluctuation $y=\delta\dot\phi/\omega_0$, which we measure as a function of an offset frequency $f$ from the carrier frequency $f_c=\Omega/(2\pi)$. From Eqs. \eqref{eqn:delta_adot}-\eqref{eqn:delta_phidot}, we readily find that
\begin{equation} 
S_y(f)=\frac{I_\phi(a_{\rm ss})}{\omega_0^2}+\left(\frac{3\gamma a_{\rm ss}}{2\omega_0^2}\right)^2\frac{I_a(a_{\rm ss})}{(2f)^2+(\Gamma_{nl}a_{\rm ss}^2)^2}.
\label{eqn:Sy}
\end{equation}

By substituting Eq. \eqref{eqn:Sy} into Eq. \eqref{eqn:allan_def} and carry out the integration, we obtain the following expression for the Allan variance
\begin{equation} 
\sigma_y^2(\tau)=\frac{I_\phi(a_{\rm ss})}{2\omega_0^2\tau}+\left(\frac{3\gamma}{4\omega_0^2\Gamma_{nl}a_{\rm ss}^2\tau}\right)^2\frac{I_a(a_{\rm ss})}{\pi\Gamma_{nl}}(2\pi\Gamma_{nl}a_{\rm ss}^2\tau+4\ee^{-\pi\Gamma_{nl}a_{\rm ss}^2\tau}-\ee^{-2\pi\Gamma_{nl}a_{\rm ss}^2\tau}-3).
\label{eqn:allan_para}
\end{equation}

Note that for large $\Gamma_{nl}a_{\rm ss}^2$, the A-F noise conversion becomes negligible. Moreover, while the Allan variance in Eq. \eqref{eqn:allan_para} is given for any integration time $\tau$, actual linearly increasing phase diffusion only occurs for integration times 
$\tau\gg\tau_c=(\Gamma_{nl}a_{\rm ss}^2)^{-1}$. In this limit, the Allan variance reduces to 
\begin{equation} 
\sigma_y^2(\tau\gg\tau_c)=\left[\frac{I_\phi(a_{\rm ss})}{2\omega_0^2}+2I_a(a_{\rm ss})\left(\frac{3\gamma}{4\omega_0^2\Gamma_{nl}a_{\rm ss}}\right)^2\right]\frac{1}{\tau}-\frac{3I_a(a_{\rm ss})}{\pi\Gamma_{nl} }\left(\frac{3\gamma}{4\omega_0^2\Gamma_{nl}a_{\rm ss}^2}\right)^2\frac{1}{ \tau^2},
\label{eqn:allan_para_ss}
\end{equation}
which reveals that $\sigma_y^2(\tau)$ can initially increase as a function of $\tau$ due to A-F noise conversion for short integration times $\tau<\tau_{\rm ex}$, and only then to decrease ($\tau>\tau_{\rm ex}$) as expected form a standard Allan variance of typical oscillator short-term stability. To observe this peculiar behavior, the condition
$$(\Gamma_{nl}a_{\rm ss}^2)^{-1}=\tau_{c}<\tau_{\rm ex}=\frac{6I_a(a_{\rm ss})}{\pi \Gamma_{nl} }\left(\frac{3\gamma}{4\omega_0^2\Gamma_{nl}a_{\rm ss}^2}\right)^2\left[\frac{I_\phi(a_{\rm ss})}{2\omega_0^2}+2I_a(a_{\rm ss})\left(\frac{3\gamma}{4\omega_0^2\Gamma_{nl}a_{\rm ss}}\right)^2\right]^{-1}$$
must be met, which we did not detect in our experiments.

\subsection{Allan variance of Duffing oscillator}

Allan variance can be calculated for a direct Duffing oscillator in the same manner as described in S1.2. From Eqs. \eqref{eqn:delta_adot_duff}-\eqref{eqn:delta_phidot_duff} we obtain the PSD of fractional-frequency fluctuation
\begin{equation}
    S_y(f) = \frac{I_\phi}{\omega_0^2} + \left(\frac{3\gamma a_{\rm ss}}{4\omega_0^2} + \frac{S\cos\Delta}{2\omega_0^2 a_{\rm ss}^2}\right)^2\frac{I_a}{f^2 + \Gamma_l^2}.\label{eq:Sy_Duff}
\end{equation}

By substituting Eq. \eqref{eq:Sy_Duff} into Eq. \eqref{eqn:allan_def}, we find the following expression for the Allan variance
\begin{equation}
    \sigma_y^2 = \frac{I_\phi(a_{\rm ss})}{2\omega_0^2\tau} + \left(\frac{3\gamma a_{\rm ss}}{8\omega_0\Gamma_l\tau} + \frac{S\cos\Delta}{4\omega_0\Gamma_la_{\rm ss}\tau}\right)^2\frac{I_a(a_{\rm ss})}{\pi\omega_0^2\Gamma_l}\left(2\pi\Gamma_l\tau + 2e^{-2\pi\Gamma_l\tau} - \frac{1}{2}e^{-4\pi\Gamma_l\tau }- \frac{3}{2}\right).
    \label{eq: Allanv_duff}
\end{equation}

For integration times $\tau\gg\tau_c = \Gamma_l^{-1}$, the Allan variance reduces to
\begin{equation}
    \sigma_y^2(\tau\gg\tau_c) = \left[\frac{I_\phi(a_{\rm ss})}{2\omega_0^2}+{2I_a}(a_{\rm ss})\left(\frac{3\gamma a_{\rm ss}}{8\omega_0^2\Gamma_l} + \frac{S\cos\Delta}{4\omega_0^2\Gamma_la_{\rm ss}}\right)^2\right]\frac{1}{\tau} - \frac{3I_a(a_{\rm ss})}{2\pi\Gamma_l}\left(\frac{3\gamma a_{\rm ss}}{8\omega_0^2\Gamma_l} + \frac{S\cos\Delta}{4\omega_0^2\Gamma_la_{\rm ss}}\right)^2\frac{1}{\tau^2}.
\end{equation}

By setting $\Delta = \pi/2$, we find
\begin{equation}
    \sigma_y^2(\tau\gg\tau_c) = \left[\frac{I_\phi(a_{\rm ss})}{2\omega_0^2}+2I_a(a_{\rm ss})\left(\frac{3\gamma a_{\rm ss}}{8\omega_0^2\Gamma_l} \right)^2\right]\frac{1}{\tau} - \frac{3I_a(a_{\rm ss})}{2\Gamma_l}\left(\frac{3\gamma a_{\rm ss}}{8\omega_0^2\Gamma_l}\right)^2\frac{1}{\tau^2}.
    \label{eq: Allan_Duff}
\end{equation}

\vspace{1em}
\begin{center}
    \refstepcounter{section}
    \label{sec:BackboneDrift}
    \large\textbf{S5. Frequency drift due to drive fluctuations}
\end{center}
\normalsize
\vspace{0.5em}

\setcounter{subsection}{0}
\setcounter{equation}{0}
\setcounter{figure}{0}
\setcounter{table}{0}

In this section, we provide a minimal explanation of the observed differences in long-term frequency stability between parametric and directly driven oscillators. For simplicity, we confine ourselves to the amplitude and frequencies associated with the backbone curve of both oscillators. We introduce slow actuation fluctuations through
\begin{equation}
S(t)=S_0+\delta S(t),\qquad \Delta(t)=\Delta_0+\delta\Delta(t),
\label{eq:SDeltaFluct}
\end{equation}
where $S$ denotes the effective drive and $\Delta$ the corresponding phase parameter in the slow-dynamic equations. The long-term drift is thus governed by slow fluctuations of the instantaneous oscillation frequency $\Omega(t)$, labeled as $\delta\Omega(t)$. We note that the steady-state solution of the parametrically driven oscillator is given by Eq.~\ref{eq:a_ss}, which leads to the following compact formula for the backbone curve.
\begin{equation}
\Omega
=
\omega_0+\frac{1}{4\omega_0}\left(\frac{3\gamma}{2 \omega_0} a_{\mathrm{ss}}^2-S\cos\Delta\right).
\label{eq:ParamBackbone}
\end{equation}

Equation~\eqref{eq:ParamBackbone} shows that the parametric backbone contains an explicit actuator-dependent term $-S\cos\Delta$. Therefore, slow pump fluctuations generate direct frequency drift:
\begin{equation}
\delta\Omega
=
-\frac{\cos\Delta_0}{4\omega_0}\,\delta S
+
\frac{S_0\sin\Delta_0}{4\omega_0}\,\delta\Delta
\qquad (\text{at fixed }a_{\mathrm{ss}}).
\label{eq:ParamDriftSensitivities}
\end{equation}
which is a direct mechanism that can degrade the long-term stability of the oscillator. We note, however, that the Duffing backbone curve depends only on the oscillation amplitude $(\Omega
=\omega_0+\frac{3\gamma}{8\omega^2_0} a^2)$ and has no explicit dependence on the drive amplitude and phase. In other words, the drive fluctuations $\delta S(t)$ and $\delta\Delta(t)$ can affect the oscillation frequency only indirectly through induced amplitude fluctuations $\delta a(t)$. Therefore,  in the parametric case, slow actuator fluctuations generate frequency drift through an \emph{additive} pathway,
$\delta S,\delta\Delta \rightarrow \delta \Omega$, while for directly driven nonlinear oscillator such fluctuations influence $\Omega$ mainly through the \emph{indirect} pathway
$\delta S,\delta\Delta \rightarrow \delta a \rightarrow \delta \Omega$.
\newpage
\setcounter{section}{6}
\vspace{1em}
\begin{center}
    \large\textbf{S6. Negligible impact of nonlinear damping in the graphene Duffing oscillator}
\end{center}
\normalsize
\vspace{0.5em}

We performed additional measurements on the $12\,\mu\mathrm{m}$ diameter bilayer graphene nanodrum as a function of drive amplitude to map the evolution of direct resonances (\autoref{fig:low-amp-peak}a). The results show that the peak amplitude of the Duffing response scales linearly with the drive strength (\autoref{fig:low-amp-peak}b), in excellent agreement with the expected relation $S = 2\omega_0 \Gamma_l a_{\rm ss}^{\rm peak}$. This confirms that nonlinear damping is negligible for direct resonances at low oscillation amplitudes as well.

\setcounter{subsection}{0}
\setcounter{equation}{0}
\setcounter{figure}{0}
\setcounter{table}{0}

\begin{figure}[H]
    \centering
    \includegraphics[width=\linewidth]{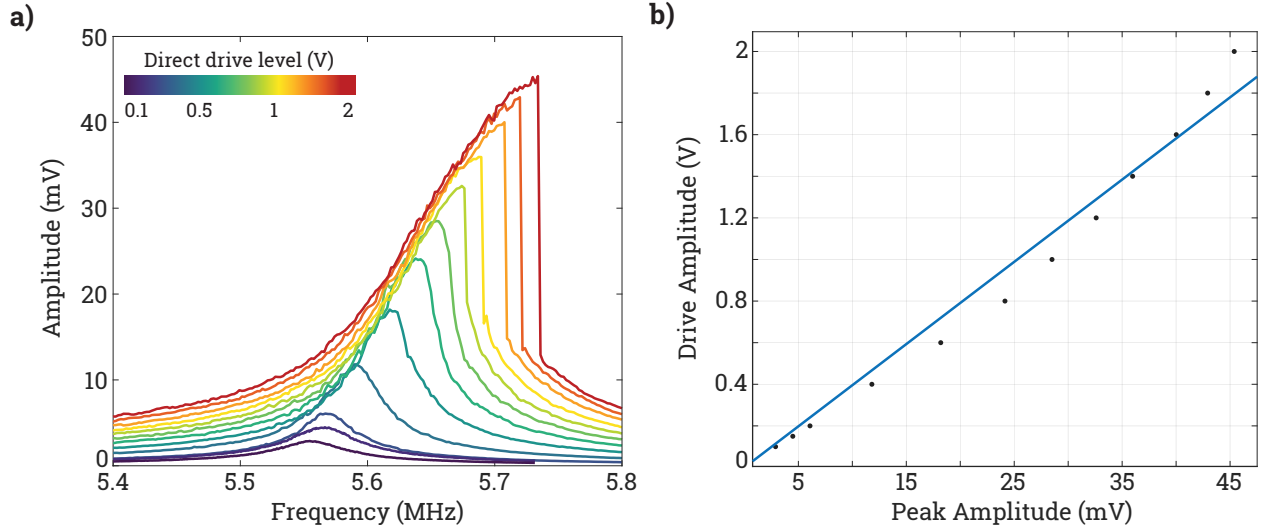}
    \caption{Drive amplitude vs peak amplitude relationships of the direct-driven case at low oscillation amplitudes.
a) Direct-driven frequency sweeps with V\textsubscript{drive} from 0.1 V to 2 V. These sweeps are taken using a $12\,\mu\mathrm{m}$ diameter bilayer graphene nanodrum.
b) Drive amplitude vs. peak amplitude relationships of the direct-driven responses shown in a). The blue line represents a linear fit to $S = 2\omega_0 \Gamma_l a_{\rm ss}^{\rm peak}$ ($\mathrm{R^2} \approx 0.98$).}
    \label{fig:low-amp-peak}
\end{figure}
